\documentclass[trackchanges,twocolumn]{aastex701}

\usepackage{multirow}
\usepackage{amsmath}
\usepackage{graphicx}
\usepackage{colortbl}
\usepackage{subcaption} 
\usepackage{dirtytalk}
\usepackage{comment}
\usepackage{soul}
\usepackage{enumitem}
\usepackage[table, svgnames, dvipsnames]{xcolor}

\begin{document}

\title{AGN Fueling and Radio Jet Evolution in the Galaxy Group NGC 5044 \\revealed by VLBA HI Absorption and Proper-Motion Radio Observations}

\author[orcid=0000-0001-5338-4472]{Francesco Ubertosi}
\affiliation{Dipartimento di Fisica e Astronomia, Università di Bologna, via Gobetti 93/2, I-40129 Bologna, Italy}
\affiliation{Istituto Nazionale di Astrofisica - Istituto di Radioastronomia (IRA), via Gobetti 101, I-40129 Bologna, Italy}
\email[show]{francesco.ubertosi2@unibo.it}  
\correspondingauthor{Francesco Ubertosi}

\author[0000-0002-5671-6900]{Ewan O'Sullivan}
\affiliation{Center for Astrophysics $|$ Harvard \& Smithsonian, 60 Garden Street, Cambridge, MA 02138, USA}
\email{eosullivan@cfa.harvard.edu}  

\author[0000-0002-4962-0740]{Gerrit Schellenberger}
\affiliation{Center for Astrophysics $|$ Harvard \& Smithsonian, 60 Garden Street, Cambridge, MA 02138, USA}
\email{gerrit.schellenberger@cfa.harvard.edu}

\author[0000-0002-9478-1682]{William Forman}
\affil{Center for Astrophysics $|$ Harvard \& Smithsonian, 60 Garden Street, Cambridge, MA 02138, USA}
\email[]{wforman@cfa.harvard.edu}

\author[0000-0001-7509-2972]{Kamlesh Rajpurohit}
\affiliation{Center for Astrophysics $|$ Harvard \& Smithsonian, 60 Garden Street, Cambridge, MA 02138, USA}
\email[]{kamlesh.rajpurohit@cfa.harvard.edu}

\author[0000-0002-1634-9886]{Simona Giacintucci}
\affiliation{Naval Research Laboratory, 4555 Overlook Avenue SW, Code 7213, Washington, DC 20375, USA}
\email[]{simona.giacintucci@nrl.navy.mil}

\author[0000-0002-8476-6307]{Tiziana Venturi}
\affil{Istituto Nazionale di Astrofisica - Istituto di Radioastronomia (IRA), via Gobetti 101, I-40129 Bologna, Italy}
\email[]{tventuri@ira.inaf.it}

\author[0009-0007-0318-2814]{Jan Vrtilek}
\affiliation{Center for Astrophysics $|$ Harvard \& Smithsonian, 60 Garden Street, Cambridge, MA 02138, USA}
\email[]{jvrtilek@cfa.harvard.edu}

\author{Laurence P. David}
\affiliation{Center for Astrophysics $|$ Harvard \& Smithsonian, 60 Garden Street, Cambridge, MA 02138, USA}
\email[]{ldavid@cfa.harvard.edu}

\author[0000-0003-3203-1613]{Preeti Kharb}
\affiliation{National Centre for Radio Astrophysics (NCRA) - Tata Institute of Fundamental Research (TIFR), 
S. P. Pune University Campus, Ganeshkhind, Pune 411007, Maharashtra, India}
\email[]{kharb@ncra.tifr.res.in}

\author[0000-0003-2206-4243]{Christine Jones}
\affil{Center for Astrophysics $|$ Harvard \& Smithsonian, 60 Garden Street, Cambridge, MA 02138, USA}
\email[]{cjones@cfa.harvard.edu}


\begin{abstract}
The role of cooling gas in triggering active galactic nucleus (AGN) feedback in the centers of galaxy groups and clusters remains a key open question. NGC~5044, the X-ray brightest galaxy group, hosts the largest known reservoir of molecular gas among cool-core groups and exhibits multiple AGN outbursts, making it an ideal system to study AGN feeding. We present new multi-frequency Very Long Baseline Array (VLBA) observations of NGC~5044 at 1.4 GHz, 4.9 GHz, and 8.4 GHz, combining continuum imaging with HI spectroscopy. At 1.4 GHz, we recovered the previously known symmetric northeast--southwest jets extending for $\sim$5.5~pc each, along with evidence for previously undetected, more extended faint emission aligned with the older, kpc-scale outbursts. Comparison of 4.9 GHz and 8.4 GHz data from 2020 and 2024 reveals clear outward proper motion of jet components, yielding an average expansion speed of $(0.10\pm0.02)\,c$ and implying a dynamical age of $\sim$180~yr for the ejection of the parsec-scale jet components. The jet width profile suggests a transition from parabolic to conical collimation at a few $\times 10^{4}$ Schwarzschild radii. We detect a narrow, redshifted HI absorption line at $+264~\mathrm{km\,s^{-1}}$ against the VLBA core, tracing a compact, cold atomic cloud within $\sim$10--20~pc of the AGN. The close velocity correspondence with previously detected CO and HI absorption features in ALMA and MeerKAT data, respectively, demonstrates that cold atomic and molecular gas coexists in infalling clouds at parsec scales. Overall, these results provide an unprecedented high angular resolution view of AGN cycling, jet growth, and feeding in a galaxy group environment.

\end{abstract}

\keywords{
\uat{active galactic nuclei}{16}, 
\uat{radio galaxies}{1343}, 
\uat{relativistic jets}{1390}, 
\uat{neutral hydrogen clouds}{1099}, 
\uat{Very long baseline interferometry (VLBI)}{1769}, 
\uat{proper motions}{1295}.
}


\section{INTRODUCTION}\label{sec:intro}

The active galactic nuclei (AGN) hosted by the central galaxies of galaxy clusters and galaxy groups play a decisive role in regulating the thermodynamics of their gaseous atmospheres (e.g., for reviews, \citealt{mcnamaranulsen2007,mcnamaranulsen2012,gitti2012,fabian2012,eckert2021,donahue2022}). The relativistic jets launched by AGN can carve X-ray cavities and drive shocks in the hot intragroup or intracluster medium (IGrM/ICM), reducing its ability to cool to lower temperatures. Nevertheless, residual cooling proceeds and leads to the condensation of warm ($\sim 10^{4}$~K) and cold ($\sim 10^{2}$~K) gas clouds in and around the central dominant galaxy. These cold gas parcels might both trigger star formation and ultimately feed the central supermassive black hole (SMBH). 
\\\indent This interplay between heating and cooling processes is well established on kpc scales. However, the properties of the gas and the jet driving mechanisms on parsec and sub-parsec scales remain poorly constrained. Absorption-line studies against bright radio cores have detected compact cold clouds at tens to hundreds of parsecs from the nucleus \citep{david2014, tremblay2016, rose2019, rose2022, rose2023}. These absorption features likely trace the final stages of gas cooling before black hole accretion. At the same time, some studies of jet dynamics on sub-parsec scales for the AGN of central galaxies exist, but they have been so far limited to massive clusters ($\geq10^{14}$~M$_{\odot}$) and/or very bright ($\sim$Jy) central radio sources (e.g., M87 in Virgo, see \citealt{asadanakamura2012,walker2018}; or 3C84 (aka NGC1275) in Perseus, see \citealt{giovannini2018}). Resolving these processes on parsec scales for lower mass systems, where cooling and heating proceed more rapidly, requires the highest angular resolution and sensitivity available.

\begin{table*}[ht!]
     \centering
     \caption{Summary of VLBA observations and continuum images employed in this work.}\label{tab:data}
     \renewcommand*{\arraystretch}{1.2}
     \addtolength{\tabcolsep}{-0.3em}
     \begin{tabular}{lccccccr}
     \hline
     
 Obs. band & Obs. Date & t$_{obs}$  &  \texttt{R} & (\emph{u,v})-cut/(\emph{u,v})-taper & $\sigma_{rms}$ & Beam FWHM, P.A. & Figure \\
 
\hline
 \multicolumn{8}{c}{{\bf Project BS283} (PI Schellenberger)} \\
\hline

  C  (4.9 GHz) & \multirow{2}{*}{Mar. 2020} & \multirow{2}{*}{5.5h}  & \multirow{2}{*}{$+2$} & \multirow{2}{*}{--} & \multirow{2}{*}{20~$\mu$Jy/beam} & \multirow{2}{*}{4.8$\times$1.9~mas, $+13^{\circ}$} & \multirow{2}{*}{Fig. \ref{fig:CXepochs}, top left} \\
  256 MHz bandwidth & & & & & & & \\
 \hline
  X  (8.4 GHz) & \multirow{2}{*}{Mar. 2020} & \multirow{2}{*}{5.5h}  & \multirow{2}{*}{$+2$} & \multirow{2}{*}{--} & \multirow{2}{*}{15~$\mu$Jy/beam} & \multirow{2}{*}{5.0$\times$2.9~mas, $-5.6^{\circ}$} & \multirow{2}{*}{Fig. \ref{fig:CXepochs}, top center} \\
  256 MHz bandwidth & & & & & & & \\

 \hline
 \multicolumn{8}{c}{{\bf Project BU038} (PI Ubertosi)} \\
 \hline

 \multirow{6}{*}{} & \multirow{6}{*}{Aug. 2024} & \multirow{6}{*}{18h} & $-2$ & $\geq$10~M$\lambda$ & 700~$\mu$Jy/beam & 9$\times$3~mas, $0.7^{\circ}$ & --\\
 L  (1.4~GHz) &  &   & $-2$ & -- & 415~$\mu$Jy/beam & 10$\times$4~mas, $0.8^{\circ}$ & -- \\
 32~MHz bandwidth &    & & $-1$ & -- & 120~$\mu$Jy/beam & 11$\times$4~mas, $0.5^{\circ}$ & Fig. \ref{fig:continuum}, top left \\
 4096 channels &    & & $+0.5$ & -- & 43~$\mu$Jy/beam & 19$\times$7~mas, $23.1^{\circ}$ & -- \\
 $\delta_{v,ch} = 1.6$~km~s$^{-1}$ &    & & $+2$ & -- & 50~$\mu$Jy/beam & 22$\times$10~mas, $9.3^{\circ}$ & Fig. \ref{fig:continuum}, bottom left \\
 & & & $+2$ & 3~M$\lambda$ taper & 50~$\mu$Jy/beam & 50$\times$29~mas, $19.4^{\circ}$ & --\\

   \hline
 \multicolumn{8}{c}{{\bf Project BU039} (PI Ubertosi)} \\
 \hline

  C  (4.9 GHz) & \multirow{2}{*}{Sep. 2024} & \multirow{2}{*}{7h}  & \multirow{2}{*}{$+2$} & \multirow{2}{*}{--} & \multirow{2}{*}{12~$\mu$Jy/beam} & \multirow{2}{*}{4.8$\times$1.9~mas, $+13^{\circ}$} & \multirow{2}{*}{Fig. \ref{fig:CXepochs}, bottom left}\\
  512 MHz bandwidth & & & & & & & \\
 \hline
 
  X  (8.4 GHz) & \multirow{2}{*}{Sep. 2024} & \multirow{2}{*}{7h} & \multirow{2}{*}{$+2$} & \multirow{2}{*}{--} & \multirow{2}{*}{30~$\mu$Jy/beam} &  \multirow{2}{*}{5.0$\times$2.9~mas, $-5.6^{\circ}$} & \multirow{2}{*}{Fig. \ref{fig:CXepochs}, bottom center}\\
  512 MHz bandwidth & & & & & & & \\

  \hline

\end{tabular}
\tablecomments{(1) Observing band and central frequency (for the L band data we also provide the channel width); (2) date of the observations; (3) total time; (4) Robust parameter used during imaging (5) (\emph{u,v})-cut or (\emph{u,v})-taper imposed during imaging (6) rms noise of the images; (7) angular resolution given by the FWHM and position angle (P.A.) of the restoring beam; (8) Figure where the corresponding image is displayed.}

 \end{table*}

In this context, NGC~5044 ($z = 0.009$, RA=13:15:23.961, DEC=$-$16:23:07.548) emerges as an interesting laboratory to study feeding and feedback. It is the X-ray brightest galaxy group in the sky (due to its intrinsically high soft X-ray luminosity and proximity to us, see e.g., \citealt{Buote_2003}), possesses the largest amount of molecular gas among cool-core groups, and shows multiple cycles of AGN outbursts \citep{gastaldello2009, david2014, schellenberger2020, rajpurohit2025, tamhane2026, temi2026}. The group has an intermediate richness that places it between small groups and galaxy clusters, with $\geq110$ spectroscopically confirmed members \citep{mendel2008}. The total virial mass of the galaxy group is $M_{500} = (9.2\pm1.6)\times10^{13}$~M$_{\odot}$ based on member galaxies \citep{mendel2008} or $M_{500} = (5.3\pm0.1)\times10^{13}$~M$_{\odot}$ from X-ray observations \citep{sadibekova2024}. The central galaxy of the group hosts a SMBH at its center; in the literature, there is not a general consensus on the mass of this SMBH. For example, \citet{david2009} measured a stellar velocity dispersion of $\sigma_{\ast} \sim 237$~km/s and found $M_{\rm BH} = 2.27\times10^{8}$~M$_{\odot}$ using the $M_{\rm BH} - \sigma_{\ast}$ relation. \citet{diniz2017} estimated $M_{\rm BH} = 1.9\pm0.9\times10^{9}$~M$_{\odot}$ assuming that all gas rotation in the inner 136~pc is dominated by a concentrated central mass. Alternatively, by modeling the spectral energy distribution from radio to sub-millimeter wavelengths, and assuming an advection-dominated accretion flow (ADAF) model for the millimeter/sub-millimeter emission, \citet{schellenberger2024} measured  $M_{\rm BH} = 2.2\times10^{9}$~M$_{\odot}$. These estimates suggest that the SMBH in NGC~5044 has a mass in the range $2\times10^{8}\lessapprox M_{\rm BH} \,\text{[M}_{\odot}\text{]}\,\lessapprox 2\times10^{9}$. Evidence of AGN feedback on kpc scales comes from the detection of multiple X-ray cavities in the IGrM \citep{gastaldello2009,david2009,david2017,schellenberger2021} and of extended, faint radio emission associated with multiple jet outbursts \citep{giacintucci2011,rajpurohit2025}. These outbursts are aligned in the northwest - southeast direction. On sub-kpc scales, Very Long Baseline Interferometry (VLBI) observations at 4.9~GHz (C band) and 8.4 GHz (X band) have revealed a parsec-scale core-jets structure with multiple knots \citep{schellenberger2021}. Interestingly, the jets on parsec scales are misaligned by nearly 90$^{\circ}$ from the kpc scale outbursts \citep{schellenberger2021,ubertosi2024a}. Extended H$\alpha$ filaments and a $4\times10^{7}$~M$_\odot$ molecular gas reservoir provide evidence for ongoing cooling despite AGN heating \citep{david2014,david2017,schellenberger2020}. Atacama Large Millimeter Array (ALMA) spectroscopy has detected narrow CO absorption features consistent with infalling molecular clouds of $\sim 7\times10^{3}$~M$_\odot$ within $\lesssim$20~pc of the nucleus \citep{schellenberger2020,schellenberger2021,schellenberger2024}. MeerKAT L-band (0.85 -- 1.67 GHz) observations found evidence of narrow HI absorption lines at similar redshifted velocites as the CO(2-1) absorption lines, suggesting that multiphase cold gas is present at the core of this elliptical galaxy \citep{rajpurohit2025}. Importantly, at a redshift of z = 0.009, we can resolve sub-parsec scales using observations at milliarcsecond angular resolution.

In this work, we present the results of a recent Very Long Baseline Array (VLBA) multifrequency campaign targeting NGC 5044 with both continuum and spectral line setups. Throughout this work, we assume a $\Lambda$CDM cosmology with $H_{0} = 70$~km~s$^{-1}$/Mpc, $\Omega_{m} = 0.3$, and $\Omega_{\Lambda} = 0.7$. Uncertainties are reported at 1$\sigma$ unless otherwise stated. The spectral index $\alpha$ is defined as $S_{\nu}\propto\nu^{\alpha}$, where $S_{\nu}$ is the flux density at frequency $\nu$. Following \citet{schellenberger2020}, we consider a systemic velocity of 2757~km~s$^{-1}$ (heliocentric reference frame) for NGC~5044 and a
luminosity distance of 31.2 Mpc \citep{tonry2001}, which
gives a linear scale conversion of $1" = 0.15$~kpc in the rest frame of NGC~5044.


\section{DATA REDUCTION}\label{sec:data}

\begin{table}[ht!]
     \centering
     \caption{Summary of VLBA cubes employed in this work for the analysis of the HI in NGC~5044.}\label{tab:datacube}\renewcommand*{\arraystretch}{1.2}
     \begin{tabular}{lccc}
     \hline     
  $\delta v$ & \texttt{R} & $\sigma_{rms,chan}$ & Beam FWHM \\
 
\hline
 
  \multirow{2}{*}{1.66~km~s$^{-1}$} & $+0.5$ & 0.9~mJy/beam & 18.6$\times$14.4~mas, $19.2^{\circ}$ \\
   &  $+2$ & 0.9~mJy/beam & 24.0$\times$10.9~mas, $5.2^{\circ}$ \\
  \cline{1-4}
   3.3~km~s$^{-1}$ &  $+0.5$ & 0.7~mJy/beam & 19.1$\times$12.5~mas, $20.8^{\circ}$  \\

  \hline
\end{tabular}
\tablecomments{(1) Velocity resolution; (2) Robustness used during imaging; (3) rms noise per channel of the cube; (4) angular resolution.}
 \end{table}

\begin{figure*}[ht!]
    \centering
    \includegraphics[width=\linewidth]{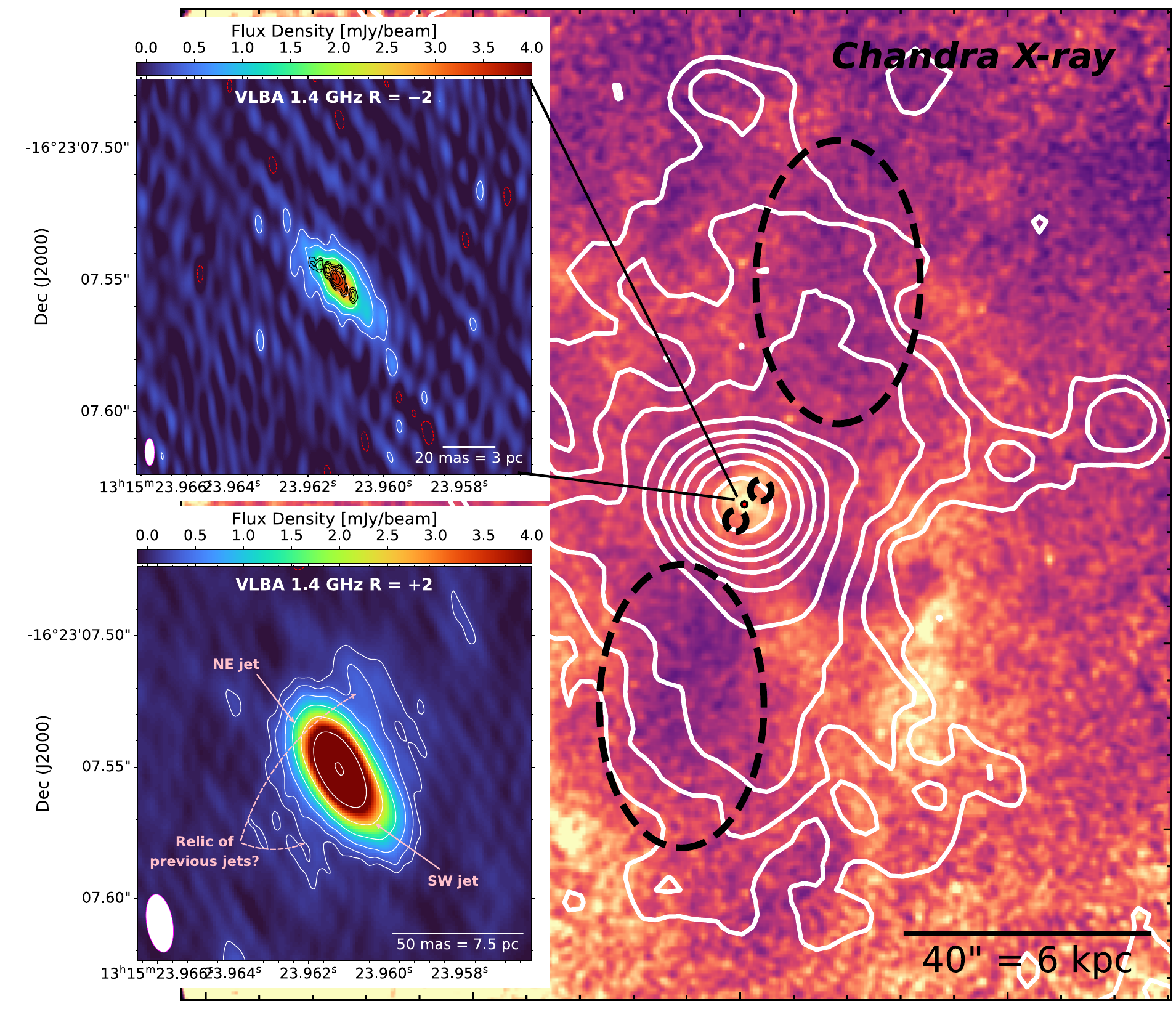}
   
    \caption{{\it Left panels}: VLBA L band continuum images of the radio core of NGC~5044 obtained with Briggs weighting with robust parameter \texttt{R=-1} (top), \texttt{R=+2} (bottom). White contours start at 3$\sigma_{rms}$ and increase by a factor of 2. In the top panel, black contours trace the 4.9~GHz emission seen in VLBA images from 2020 \citep{schellenberger2021}. {\it Main background panel:} {\it Chandra} X-ray residual image of NGC~5044 from \citet{ubertosi2024a}, showing the hot IGrM. We overlay in black the dashed ellipses corresponding to the AGN-driven bubbles in the gas (see \citealt{david2017,schellenberger2021,ubertosi2024a,rajpurohit2025}). We also overlay in white the contours from the uGMRT Band 3 image at $10''$ resolution of NGC~5044 presented in \citet{rajpurohit2025}. The contours are drawn at 3, 6, 12...$\times\sigma_{rms}$, where $\sigma_{rms}=21\mu$Jy~beam$^{-1}$ is the rms noise of the radio image. The central red circle shows the position of the AGN. }
    \label{fig:continuum}
\end{figure*}

\begin{figure}[ht!]
    \centering
    \includegraphics[width=\linewidth]{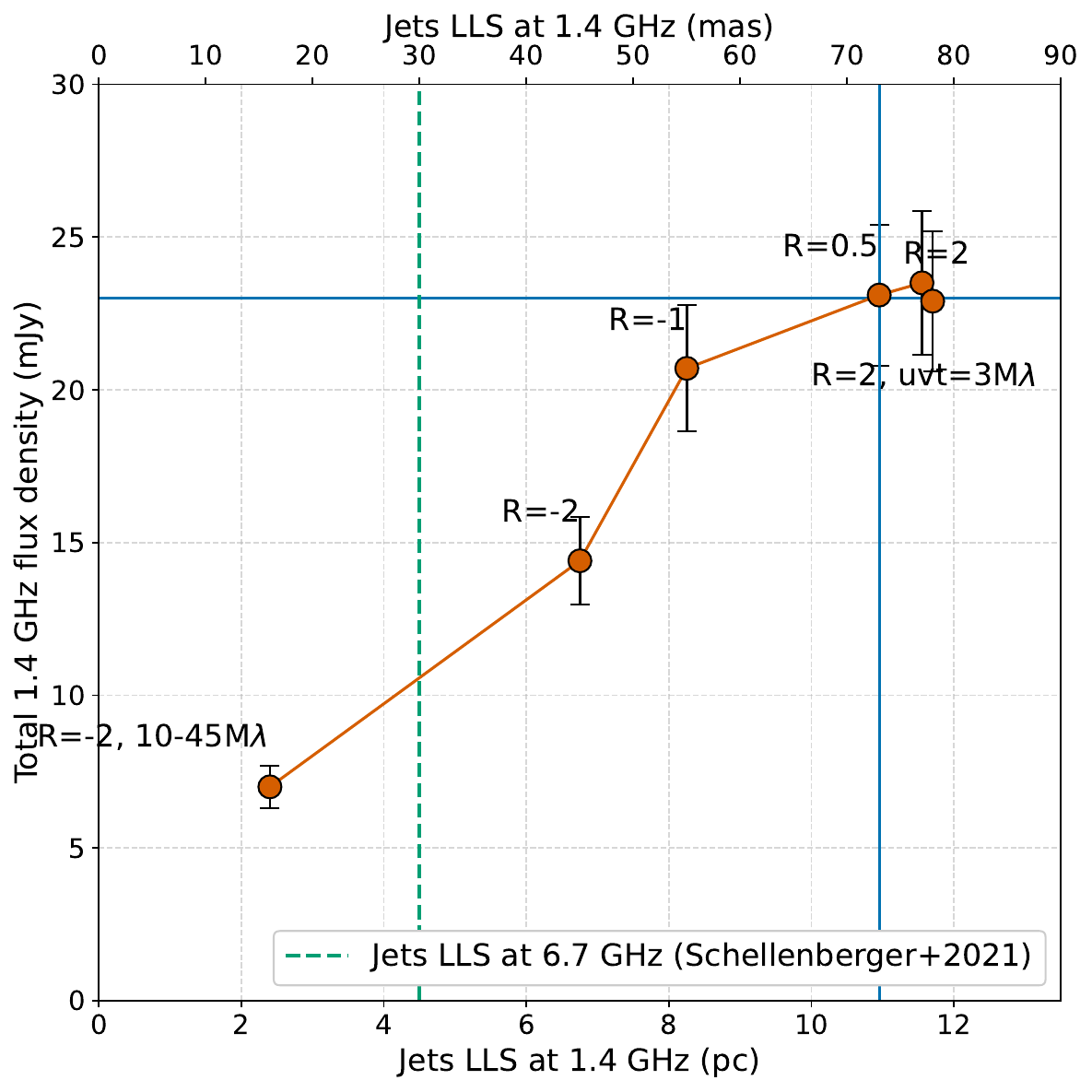}
    \caption{Total 1.4~GHz flux density as a function of Briggs weighting (and optional tapering or (\emph{u,v})-selection) used during imaging, plotted as a function of the largest linear size of the 5$\sigma_{rms}$ contour in the final image. The flux density saturates at $\sim$23~mJy for \texttt{R~$\geq+$0.5}, suggesting that the largest linear size of the jets in NGC~5044 is $\sim$11~pc.}
    \label{fig:LLS-jets}
\end{figure}

We analyze phase-referenced VLBA observations of NGC~5044 (see Table \ref{tab:data} for full details) from March 2020 (C, X bands, 11h, project BS~283, PI Schellenberger), August 2024 (L band, 18h, project BU~038, PI Ubertosi), and September 2024 (C, X bands, 14h, project BU~039, PI Ubertosi). Projects BS~283 and BU~039 were divided into two observing sessions at each frequency, while project BU~038 was divided into six observing sessions. Below, we describe the calibration of the C \& X band data and the L-band data separately.

\paragraph{C \& X bands (continuum)} The data from projects BS~283 and BU~039 include observations of NGC~5044 at C band (4.9 GHz) \& X band (8.4 GHz). The observations from project BS~283 were taken with 256 MHz bandwidth divided into 512 channels, while the observations from project BU~039 were taken with 512 MHz bandwidth divided into 1024 channels. The data were reduced using standard techniques in AIPS \citep{aips1990}, including correction of Earth orientation parameters, ionospheric delays, instrumental delays on phases, and bandpass. We fringe-fitted the data and transferred the solution from the phase calibrator (J1317-1345\footnote{See \url{https://obs.vlba.nrao.edu/cst/calibsource/10819}.} both for project BS~283 and for project BU~039) to the target. Ultimately, we applied the calibration to the data, averaged channels within the different spectral windows, and split out a corrected dataset. 
We exported the calibrated (\emph{u,v})-data from AIPS and used the software CASA \citep{casa2022} for self-calibration. To properly register the two epochs, we used the image obtained from the first cycle of self-calibration of the 2020 data as a starting model for the self calibration of the 2024 data (e.g., \citealt{giroletti2003}). Five cycles of self-calibration were performed (four phase-only + one amplitude and phase), after which the final rms noise (Table \ref{tab:data}) at C and X band is consistent with the theoretical rms noise expected from the data. Final images were produced using a \texttt{multiscale} deconvolution algorithm (with scales of 0, 1, 4 beams), selecting Briggs weighting \citep{briggs1995} with robust parameter \texttt{R=2} (see Table \ref{tab:data}). We attempted joint imaging of the 2020 and 2024 datasets at matching frequency to produce a higher sensitivity continuum image, but due to proper motion of jet components and relatively strong variability of the core flux density (see Section \ref{sec:results}), the resulting images are characterized by a relatively low fidelity. 

\paragraph{L band (continuum \& spectral line)} We observed NGC~5044 at L band with 32 MHz bandwidth divided into 4096 channels, giving a channel width of $\Delta\nu = 7.8$~kHz (1.66~km~s$^{-1}$). The band was centered at 1.4074~GHz (1.3914~GHz -- 1.4234~GHz), which matches the redshifted frequency of the HI 21-cm line (rest frequency 1.42041~GHz) at the systemic velocity of NGC~5044 ($v = 2757$~km~s$^{-1}$; \citealt{schellenberger2020}). The phase calibrator was  J1317-1345\footnote{See \url{https://obs.vlba.nrao.edu/cst/calibsource/10819}.} for this project. We reduced the data in AIPS using the same approach adopted for the C and X band data until the final self-calibration part. For this, starting from the phase-referenced (\emph{u,v})-data we produced a “continuum" dataset by averaging the visibilities by a factor of 10 in frequency (velocity resolution 16.6~km~s$^{-1}$), to increase the signal to noise ratio. This continuum dataset was then used for self-calibration in CASA (5 cycles of phase-only self-calibration). From the final self-calibrated continuum file we obtained final continuum images of NGC~5044, using a \texttt{multiscale} deconvolution algorithm (with scales of 0, 1, 4 beams), and Briggs weighting scheme \citep{briggs1995}. To explore the variations of the total recovered flux density or image structures with the angular resolution and sensitivity of the images, we adopted various robust parameters (from \texttt{R}$=-$2, i.e., uniform weighting, to \texttt{R}$+$2, i.e., natural weighting) and optional inner (\emph{u,v})-cut (to increase the angular resolution) or (\emph{u,v})-taper (to maximize the sensitivity). The different combinations are reported in Table \ref{tab:data}). We transferred the self-calibration solutions from the continuum data to the line data of the target (at the original velocity resolution of 1.66~km~s$^{-1}$). The spectral-line data were Doppler corrected to a Kinematic Local Standard of Rest (LSRK) reference frame (to facilitate comparison with the ALMA data presented in \citealt{schellenberger2020}). We subtracted the continuum model with the task \texttt{uvsub} and fitted a first order polynomial (a line) with the task \texttt{uvcontsub} to remove any residual continuum emission. Finally, spectral-line cubes between 1.404~GHz and 1.410~GHz were produced at the native frequency resolution using Briggs weighting with robust parameters of $+0.5$ and $+2$. Additional cubes were generated after binning the data by a factor of two in frequency, with robust $=+0.5$ (see Table~\ref{tab:datacube} for details).\\
\par When measuring flux densities $S_{\nu}$ from the images at frequency $\nu$ (1.4 GHz, 4.9 GHz, or 8.4 GHz), we computed the uncertainties $\delta S_{\nu}$ as:
\begin{equation}\label{eq:uncertflux}
    \delta S_{\nu} = \sqrt{N_{b}\sigma_{\text{rms}}^{2} + (f_{\%}S_{\nu})^{2}},
\end{equation}
where $N_{b}$ is the ratio between the area within which the flux density is computed and the beam area, $\sigma_{\text{rms}}$ is the rms noise, and $f_{\%}$ is the uncertainty on the flux density scale (5\% for the VLBA).

\begin{figure*}[ht!]
    \centering
    \includegraphics[width=0.535\linewidth]{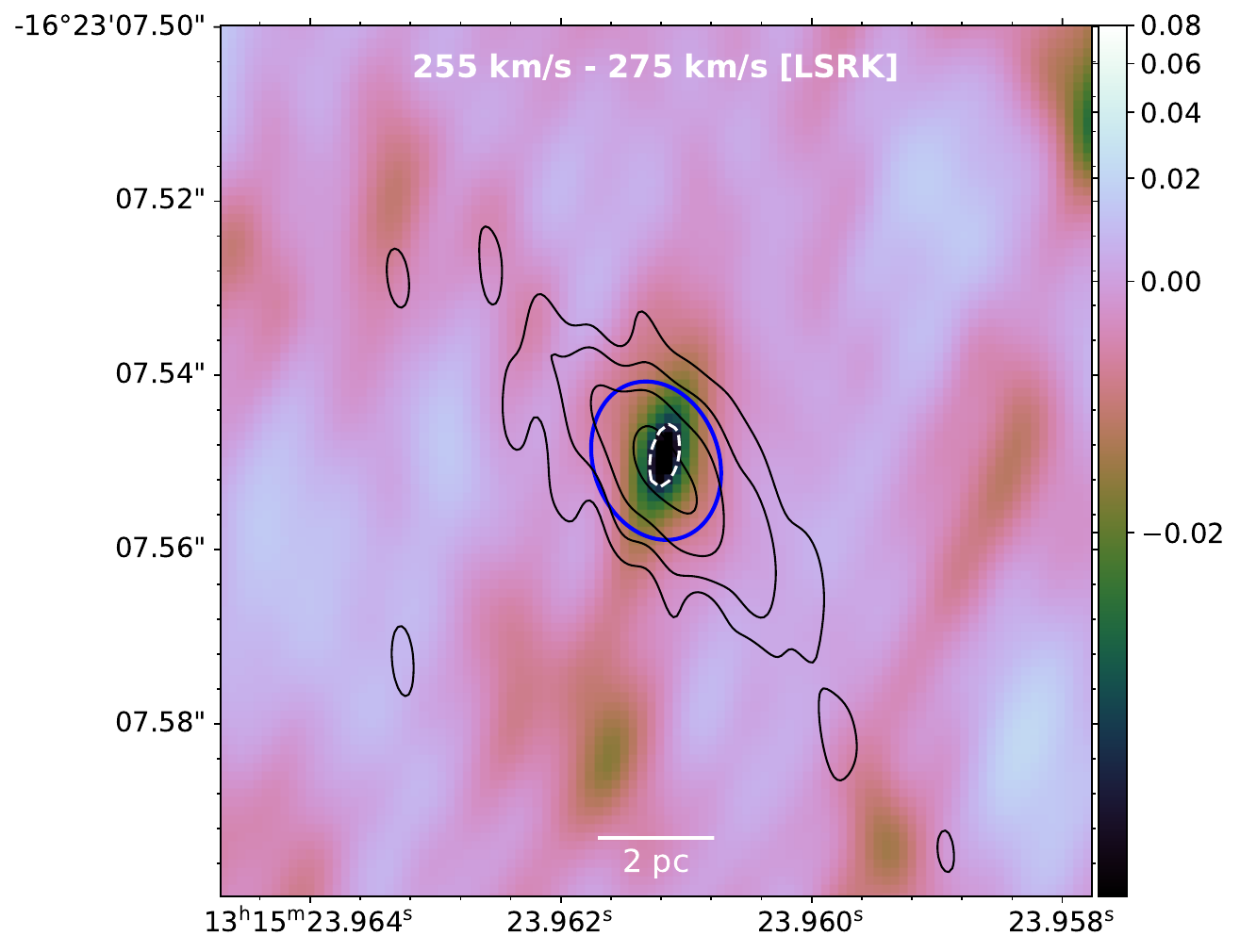}
    \includegraphics[width=0.44\linewidth]{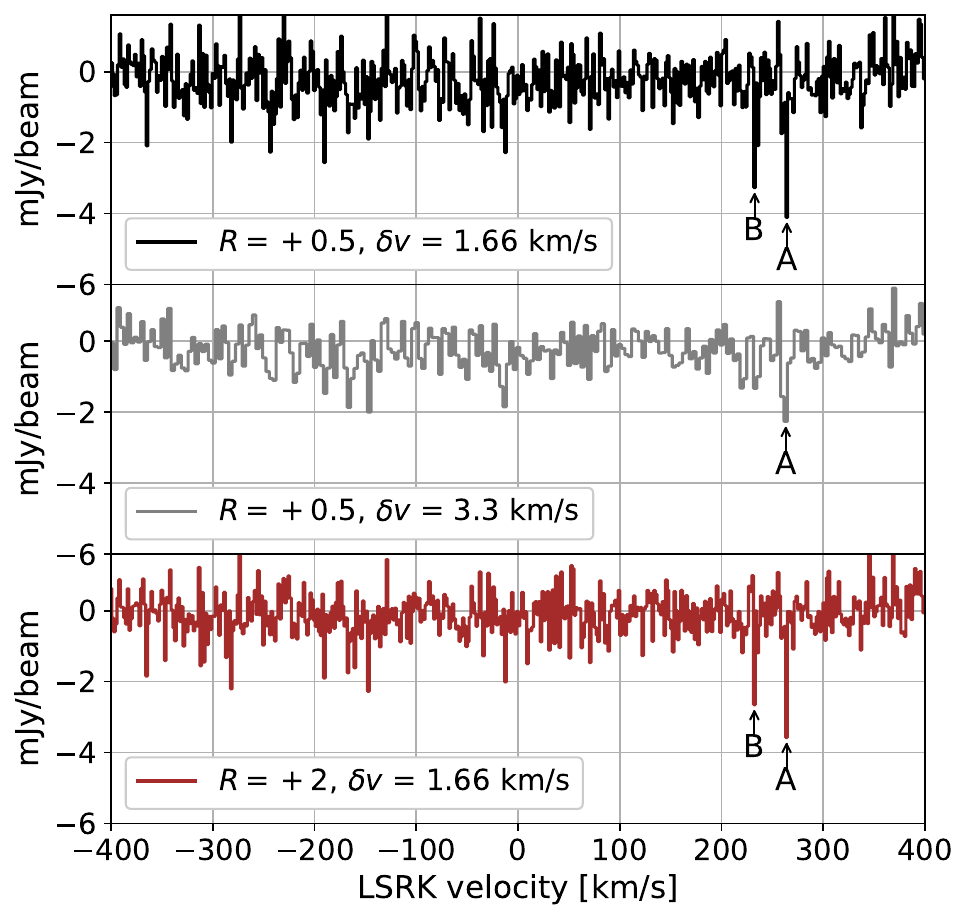}
    \caption{Atomic HI absorption against the radio core in NGC~5044. {\it Left panel:} moment 0 map obtained by integrating the signal between $v = 255$~km~s$^{-1}$ and $v = 275$~km~s$^{-1}$ in the \texttt{R$=+0.5$}, $\delta v = 1.66$~km~s$^{-1}$ cube, showing a $\sim5\sigma$ (white contour) absorption feature coincident with the radio core of NGC~5044. Black contours trace the continuum emission from the image shown in Figure~\ref{fig:continuum} (top left), while the blue ellipse shows the beam-sized region used to extract the spectra. {\it Right:} HI spectra against the radio core of NGC~5044 extracted from three data cubes, see Table~\ref{tab:datacube} for imaging details. The narrow HI absorption lines, identified as significant and discussed in the text, are labeled. We note that the three spectra were extracted from the same data but imaged using different channel resolution or robustness.}
    \label{fig:HIcore}
\end{figure*}

\section{RESULTS}\label{sec:results}

\subsection{L band observations}
\subsubsection{Continuum analysis}\label{subsec:1.4continuum}
We show in Figure~\ref{fig:continuum} (top left) the VLBA 1.4 GHz continuum image obtained with Briggs weighting with robust parameter \texttt{R=-1}, which reveals a bright central region (the core) surrounded by nearly symmetric extensions (jets) in the northeast - southwest direction. We also overlay the 4.9 GHz contours from the 2020 VLBA dataset \citep{schellenberger2021}, after matching the position of the brightest pixels in the two images.  
It is clear that while the orientation of the jets is very similar, the jets appear more extended at 1.4 GHz than at 4.9 GHz, likely due to a combination of higher sensitivity to extended emission of the L band data and of a steep spectral index of the jets. In the bottom left panel of Figure~\ref{fig:continuum}, we show the VLBA 1.4 GHz continuum image obtained with \texttt{R = +2}, which recovers the central core and the northeast-southwest jets, as well as excess radio emission (at $3\times, 6\times\sigma_{rms}$) in the perpendicular direction to the jets. These extended structures might represent the fossil remnants of the jets that excavated the kpc-scale X-ray cavities (Figure~\ref{fig:continuum}; see also \citealt{schellenberger2021,rajpurohit2025}), given the very good alignment between the X-ray cavities and the pc-scale extended radio emission. This might support the idea that NGC~5044 experienced multiple, misaligned jet episodes with a very high AGN duty cycle (as originally proposed by \citealt{schellenberger2021}), as observed in several radio galaxies \citep[e.g.,][]{Kharb_2006,Rao_2023,ubertosi2024a}. 
While interesting, we do not discuss these features in further detail, as the relatively poor (\emph{u,v})-coverage at short spacings limits a robust study of the morphology, or a definite claim of detection, of these extended components.
\par In order to understand how much extended flux density is recovered by the VLBA data, and how this relates to the maximum length of the northeast-southwest jets, we investigated how the total recovered flux density and largest linear size (LLS) of radio emission varies with the weighting used during deconvolution and imaging. Specifically, we considered the full set of L band maps listed in Table~\ref{tab:data} and measured the total flux density above 5$\sigma_{rms}$ as a function of the largest linear size of the 5$\sigma_{rms}$ contours of the image. Figure~\ref{fig:LLS-jets} shows that the total flux density increases from the $\sim$7~mJy obtained from the \texttt{R = -2}, (\emph{u,v})$\geq10$~M$\lambda$ image (recovering the unresolved core flux density only) up to $\sim$23~mJy for \texttt{R$\geq$+0.5}, after which it flattens; the use of more natural weighting or of (\emph{u,v})-tapering does not lead to recovering additional significant flux density. The LLS of the jets does not increase beyond $\sim11$~pc either. This suggests that the emission associated with the northeast - southwest outburst, while being more extended than the 4.9~GHz data originally suggested, is still confined to the pc scales and thus very recent. 
\par We adopt the total flux density measured from the map of Figure~\ref{fig:continuum} (bottom left) as representative of the total 1.4 GHz VLBA flux density of NGC~5044, i.e., $S_{1.4} = 23.5\pm2.5$~mJy. This is $\sim$4~mJy below the arcsec-scale VLA flux density at the same frequency ($27.2\pm1.1$~mJy from 1992 data, \citealt{Grossova2022}, or $27.5\pm0.5$~mJy from 2015 data, \citealt{schellenberger2021}); $\sim$4.5~mJy below the 1.4 GHz flux density of the radio core measured from MeerKAT at the highest possible resolution ($\sim$28~mJy at $4\arcsec$ from 2021, \citealt{rajpurohit2025}); and $\sim$4~mJy below the flux density measured in the The Rapid ASKAP Continuum Survey (RACS) at 1.368~GHz ($27.8\pm1.7$~mJy, \citealt{duchesne2024}). This may suggest that there is additional extended emission between the VLBA and the VLA scales which, if present, might represent a continuation of the fossil northwest-southeast outburst seen on kpc scales and tentatively in the \texttt{R = +2} VLBA image. Alternatively, as these measurements were taken at different epochs (the VLBA observations are from 2024), it is possible that time variability (reported at the 20\% level in the sub-millimeter band by \citealt{schellenberger2024}) might explain the difference in flux density. Future observations are needed to reveal the nature of this discrepancy.

\begin{figure*}[ht!]
    \centering
    \includegraphics[width=0.9\linewidth]{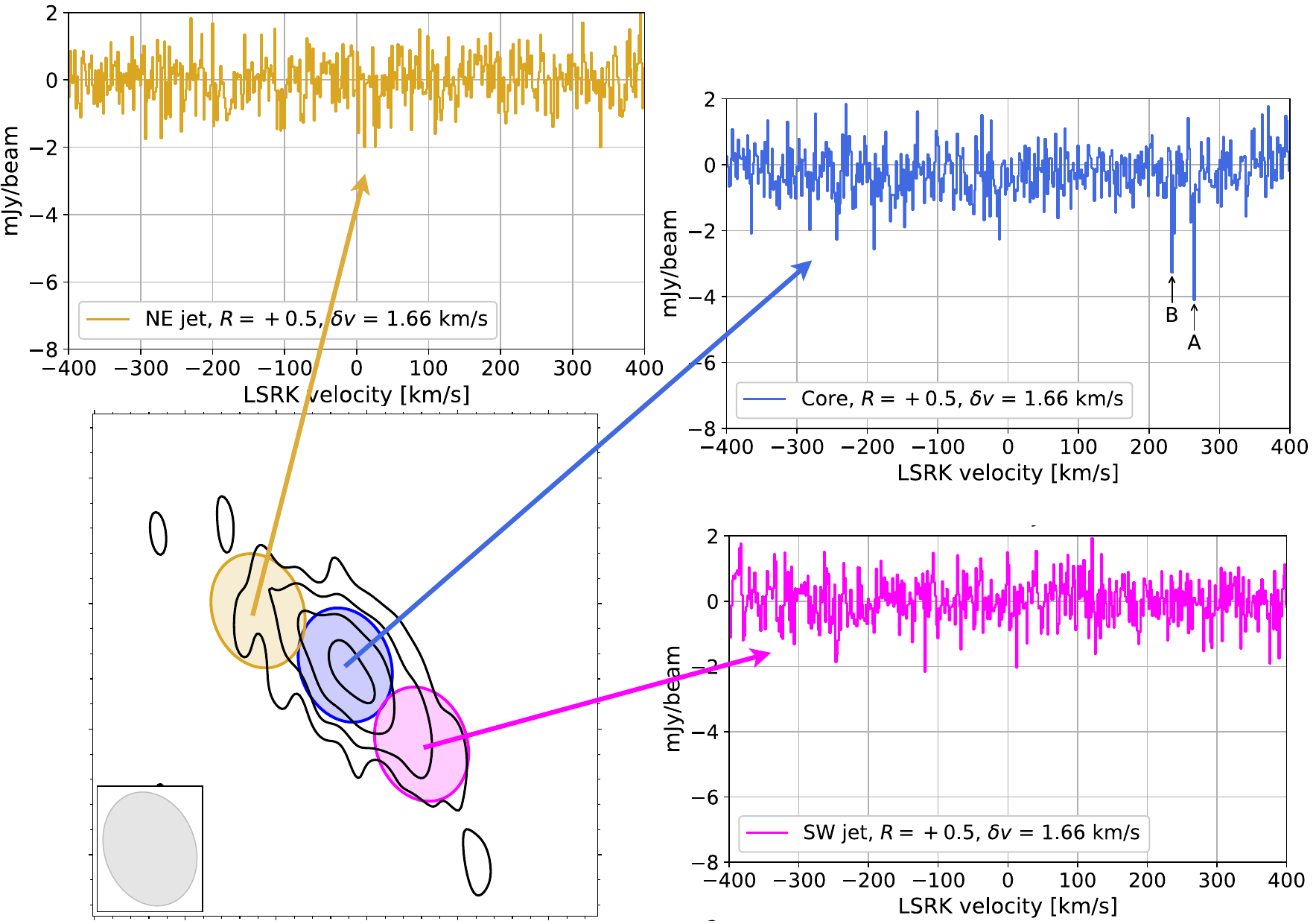}
    \caption{HI absorption against the radio core of NGC~5044. {\it Central panel:} black contours trace the continuum emission from the image shown in Figure~\ref{fig:continuum} (top left), while the ellipses show the beam-sized region used to extract the spectra. The beam is shown in the bottom-left corner. {\it Left, Top, Right panels:} HI spectra against the northeast jet, the core, and the southwest jet of NGC~5044, respectively, extracted from the data cube obtained by setting \texttt{R$=+0.5$} during imaging and with a channel width $\delta v = 1.66$~km~s$^{-1}$ (see Table~\ref{tab:datacube}). The narrow HI absorption lines, identified as significant and discussed in the text, are labeled.}
    \label{fig:HIspectra}
\end{figure*}

\subsubsection{Spectral line analysis}\label{subsec:spectrallineanalysis}

CO observations with ALMA \citep{schellenberger2020} and HI observations with MeerKAT \citep{rajpurohit2025} previously revealed the presence of molecular and atomic gas clouds against the core of NGC~5044. The ALMA detection suggested that the clouds are closer than $\sim$20~pc to the radio core, and the correspondence between the velocity of CO and HI points to the gas phases being connected. Here we investigate if the HI clouds are visible in the VLBA data, that have a superior angular resolution and higher velocity resolution relative to MeerKAT. \\
\par We analyzed the VLBA HI cubes listed in Table~\ref{tab:datacube}, focusing on the velocity range $-400$~km~s$^{-1}$ to $+400$~km~s$^{-1}$ (similarly to \citealt{david2014}; this range corresponds to $\sim$3$\times$ the stellar velocity dispersion of NGC~5044) relative to the systemic velocity of the galaxy (2757~km~s$^{-1}$). Since the continuum source is resolved (see Figure~\ref{fig:continuum}), we separately investigated the HI spectra extracted from the core and along the jets in beam-sized regions. 
\paragraph{The core spectrum} We find strong evidence for HI absorption lines against the radio core. We show in Figure~\ref{fig:HIcore} the 1.4 GHz continuum contours (from the \texttt{R $=+0.5$} image) overlaid on top of a moment 0 map. This map was obtained by integrating the \texttt{R $=+0.5$}, $\delta v = 1.66$~km~s$^{-1}$ cube between 255~km~s$^{-1}$ and 275~km~s$^{-1}$, where the CO and HI absorption lines were found by \citealt{schellenberger2020} and \citet{rajpurohit2025}, respectively. At the position of the core, a clear absorption signature is detected at $5\sigma_{rms}$ (white dashed contour in Figure~\ref{fig:HIcore}, left). Using a beam sized region centered on the core (blue ellipse), we extracted a spectrum from each cube listed in Table~\ref{tab:datacube}, with the results shown in Figure~\ref{fig:HIcore} (right panel; note that the spectra were extracted from the same data but imaged using different channel resolution or robustness). No significant HI emission features are observed; instead, narrow, redshifted absorption features are visible between 200~km~s$^{-1}$ and 300~km~s$^{-1}$. The most significant and reliable feature is line “A" at 264~km~s$^{-1}$, that is detected at more than 5$\sigma$ in the top and bottom spectra and at more than 4$\sigma$ in the middle spectrum. A potential additional line “B" at 232~km~s$^{-1}$ is visible at $\geq4\sigma$ in the top and bottom spectra only (different velocity resolutions with robustness $R=+0.5$). We report their properties in Table~\ref{tab:HIlines}, measured from the \texttt{R$=+0.5$}, $\delta v = 1.6$~km~s$^{-1}$ cube.\\
\begin{center}
\begin{table*}[ht!]
     \caption{Properties of the HI lines detected with the VLBA against the core of NGC~5044 (see Figure~\ref{fig:HIcore} and Figure~\ref{fig:HIspectra}). }\label{tab:HIlines}\renewcommand*{\arraystretch}{1.2}
     \begin{tabular}{lccccccccc}
     \hline     
  & Line & $v$ & SNR & FWHM & $S_{\nu}^{l,p}$ & $S_{\nu}^c$ & $\tau$ & $\int \tau dv$ & $N_{\text{HI,abs}}$\\
  &  & km~s$^{-1}$ &  & km~s$^{-1}$ & mJy/beam & mJy &  & km~s$^{-1}$ & cm$^{-2}$\\

\hline

 \multirow{2}{*}{Core}& A & 264.2  & 5.4 (detection) & 1.66  & $-4.1$  & 8.4  & 0.7 & 1.11 & $\leq 1.9 \times10^{21}$\\

   & B & 232.5  & 4.3 (candidate) & 1.66 & $-3.3$  & 8.4  & 0.5 & 0.82 & $\leq 1.5 \times10^{21}$ \\
  \hline

\end{tabular}
\tablecomments{(1) Extraction region (2) Line identifier; (3) velocity of the line; (4) significance of the detection; (5) width of the line; (6) peak flux density of the line; (7) continuum flux density in the extraction region; (8) peak optical depth of the absorption feature (assuming a covering factor $c_{f} = 1$); (9) integrated optical depth of the absorption feature, from Equation~\ref{eq:opticaldepth}; (10) column density of the HI absorbing gas, from Equation~\ref{eq:columndensity}.}
 \end{table*}
\end{center}
\begin{figure}[ht!]
    \centering
    \includegraphics[width=\linewidth]{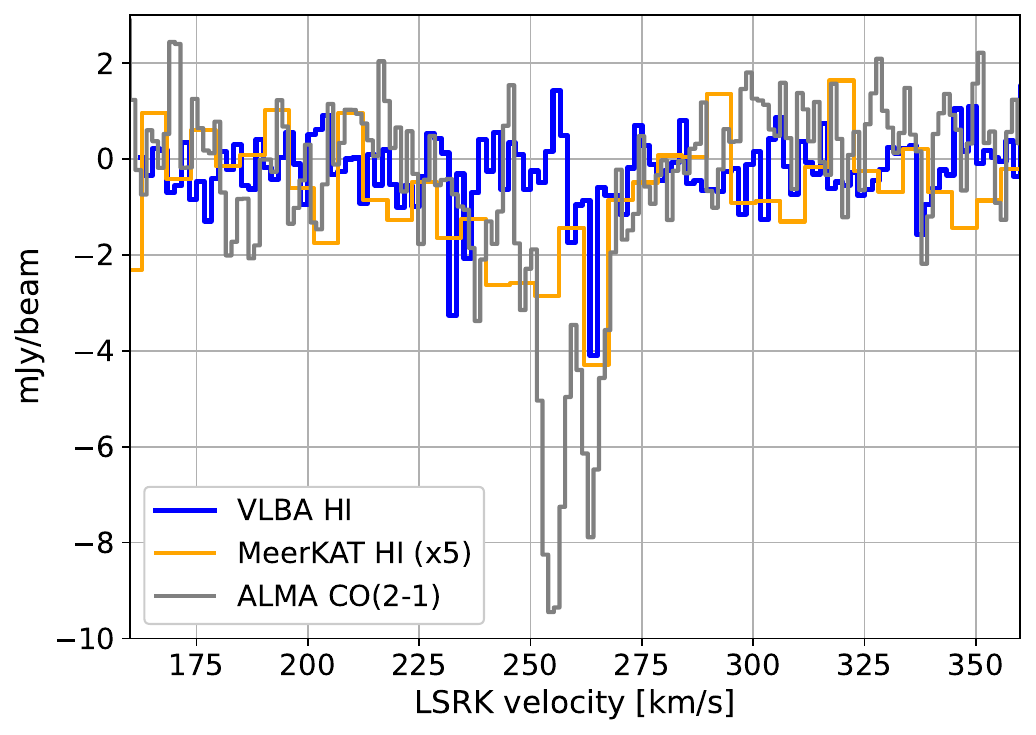}
    \caption{Atomic and molecular gas absorption features against the radio core in NGC~5044. We show the spectra of HI from VLBA data (blue; from Figure~\ref{fig:HIcore}, right panel, top spectrum) and MeerKAT data (orange and scaled by a factor of $\times$5; \citealt{rajpurohit2025}), and of CO(2-1) from ALMA data (gray; \citealt{schellenberger2020}). The spectra are zoomed on the velocity range of the redshifted atomic and molecular clouds seen in absorption.}
    \label{fig:HIandCO}
\end{figure}

\paragraph{The jet spectra} We show in Figure~\ref{fig:HIspectra} (central panel) the beam-sized regions used to extract the HI spectra against the northeast (gold ellipse) and southwest (magenta) jets. The corresponding spectra, extracted from the data cube with \texttt{R$=+0.5$}, $\delta v = 1.66$~km~s$^{-1}$ (see Table \ref{tab:datacube}), are shown in the left and right panels, respectively. For comparison, we also show the core extraction region and spectrum in blue (top panel). No significant emission or absorption features are observed in the spectrum against the jets. 
Similar results are obtained from the analysis of the other HI cubes used in the analysis. 
\\\par As a secondary check of this result, a blind search for lines in the HI spectrum against the core and jets performed in CARTA \citep{carta2021} identifies line A from the core only, and provides a best-fit FWHM of $2.11\pm0.45$~km~s$^{-1}$, consistent with the channel width of 1.66~km~s$^{-1}$. We thus adopt the channel width as a measure of the FWHM of the lines, and we consider line “A" in the core to be a strong detection, while line “B" in the core is a candidate detection.
\par We can estimate the integrated optical depth of the HI absorption, defined as:
\begin{equation}
\label{eq:opticaldepth}
    \int \tau dv = \sum_{i=1}^{n} - \ln\left(1-\frac{f_{i}}{c_{f}\times S^{c}_{\nu}}\right)\times\delta v\,,
\end{equation}
where $\tau$ is the optical depth, $f_i$ is the HI absorption flux density at frequency channel i, n is the number of channels where absorption is seen, $c_{f}$ is the covering factor of the absorbing gas relative to the continuum source, $S^c_\nu$ is the continuum flux density, and $\delta v$ is the width of each channel in km~s$^{-1}$. 
The continuum flux density in our beam-sized extraction region is $S_{\nu}^c = 8.4\pm0.9$~mJy for the core. 
We assume a covering factor $c_{f} = 1$, which returns integrated optical depths of 1.11~km~s$^{-1}$ and 0.82~km~s$^{-1}$ for line A and B in the core, respectively. 
The lines are relatively optically thick, with $S_{\nu}^{l,p}\sim50\%S_{\nu}^c/c_{f}$ for line A and $S_{\nu}^{l,p}\sim40\%S_{\nu}^c/c_{f}$ for line B (where $S_{\nu}^{l,p}$ is the peak flux density of the line) in the core spectrum. We note that, since the radio continuum source is clearly resolved (see Figure~\ref{fig:continuum}), the covering factor might be lower than unity, thus increasing the integrated optical depth. 
\\\indent From the integrated optical depth we can derive the absorbing column density of HI:
\begin{equation}
\label{eq:columndensity}
    N_{\text{HI,abs}} = 1.82\times10^{18}(T_{spin})\int\tau dv\quad[\text{cm}^{-2}]\,,
\end{equation}
where $T_{spin}$ [K] is the spin temperature representing the weighted harmonic mean of the various thermal components of HI and $dv$ has units of km~s$^{-1}$. Adopting the upper limit $T_{spin}\leq950$~K from \citet{rajpurohit2025}, we find an upper limit of $N_{\text{HI,abs}}\lessapprox 2 \times10^{21}$~cm$^{-2}$ for the line at 264~km~s$^{-1}$ (see Table~\ref{tab:HIlines}). \\
\par In Figure~\ref{fig:HIandCO}, we compare the VLBI HI (blue), MeerKAT HI (orange, scaled by $\times$5) and ALMA CO(2-1) (gray) spectra against the radio core of NGC~5044. We stress that these spectra were extracted from observations with very different angular resolution ($\sim$0.01$"$, VLBA; $\sim$8$"$, MeerKAT; $\sim$0.05$"$, ALMA). Clearly, the HI absorption feature at 264~km~s$^{-1}$ seen in VLBA has a counterpart in CO(2–1) from ALMA and in HI from MeerKAT, and is much narrower (from FWHM$\approx$7~km~s$^{-1}$ in MeerKAT and ALMA to FWHM$\approx$1.7~km~s$^{-1}$ in VLBA). The second line detected from ALMA and MeerKAT at 258~km~s$^{-1}$ (see \citealt{schellenberger2021,rajpurohit2025}) is not detected in the VLBA spectrum. We note here that while the absorption feature at $\sim$258 km s$^{-1}$ is detected in both ALMA and MeerKAT data, the MeerKAT HI line is broader by a factor of 2 and exhibits a small offset (of $6.0\pm2.4$~km~s$^{-1}$) in peak velocity relative to the narrower ALMA CO line (see Figure 8 and Table 5 in \citealt{rajpurohit2025}). This indicates that, for this second line, the structure of neutral gas along the line of sight might be different compared to the molecular gas. Line “B" of the VLBA HI spectrum lies at the lower edge of this broad absorption region in the MeerKAT spectrum (see Figure~\ref{fig:HIandCO}). 
\\\indent The fact that line “A" appears in both atomic and molecular data, while the second CO line at 258~km~s$^{-1}$ does not, is not surprising. Rather, it is likely caused by the combination of the different angular resolutions of the instruments, the presence of clouds with a range of properties (optical depth, size, composition, and distance from the source), and the frequency-dependent structure of the continuum source (see more in Section~\ref{subsec:agnfeeding}; see also \citealt{rose2023}).

\begin{figure*}[ht!]
    \centering
    \includegraphics[width=0.32\linewidth]{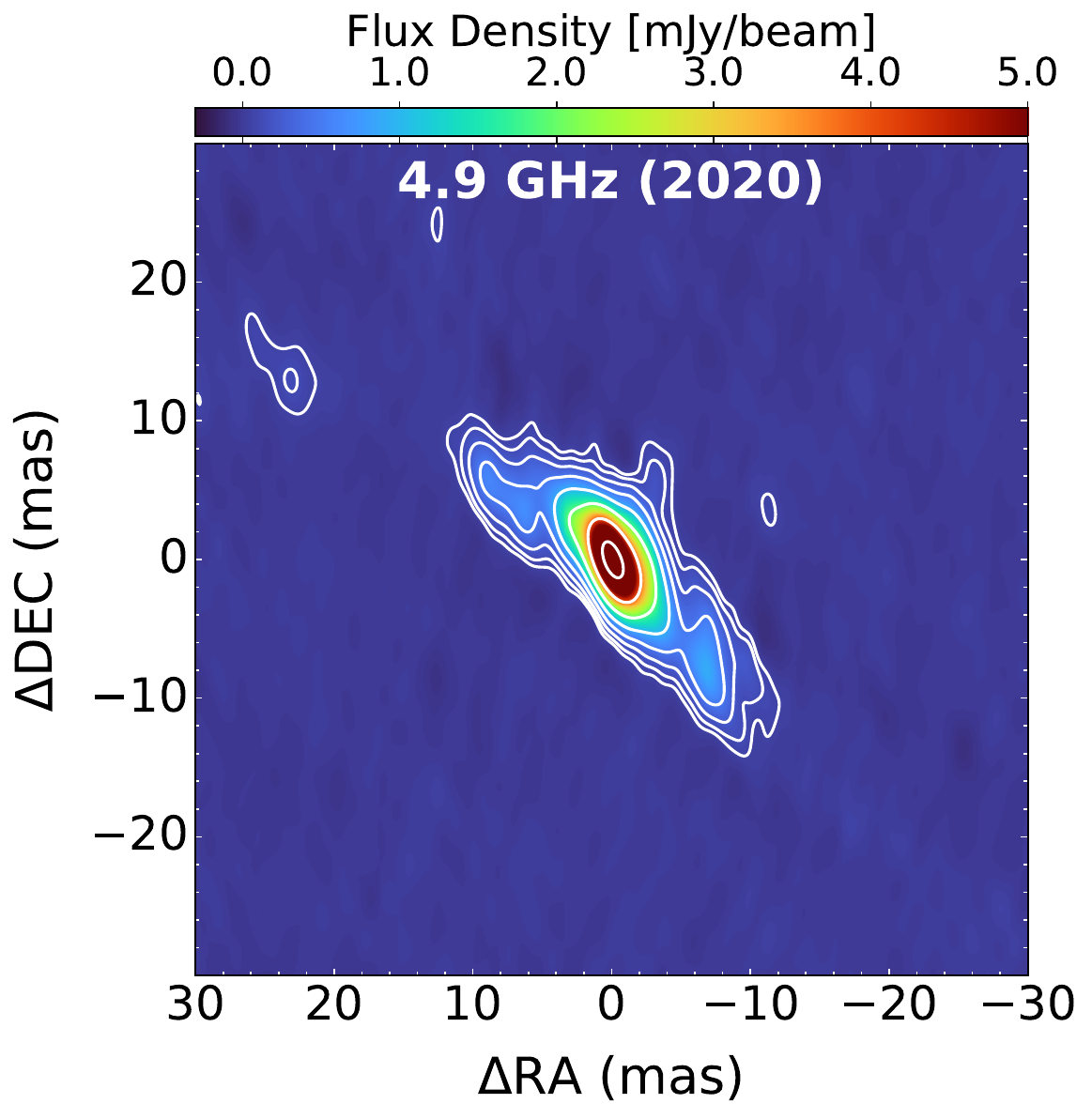}
    \includegraphics[width=0.32\linewidth]{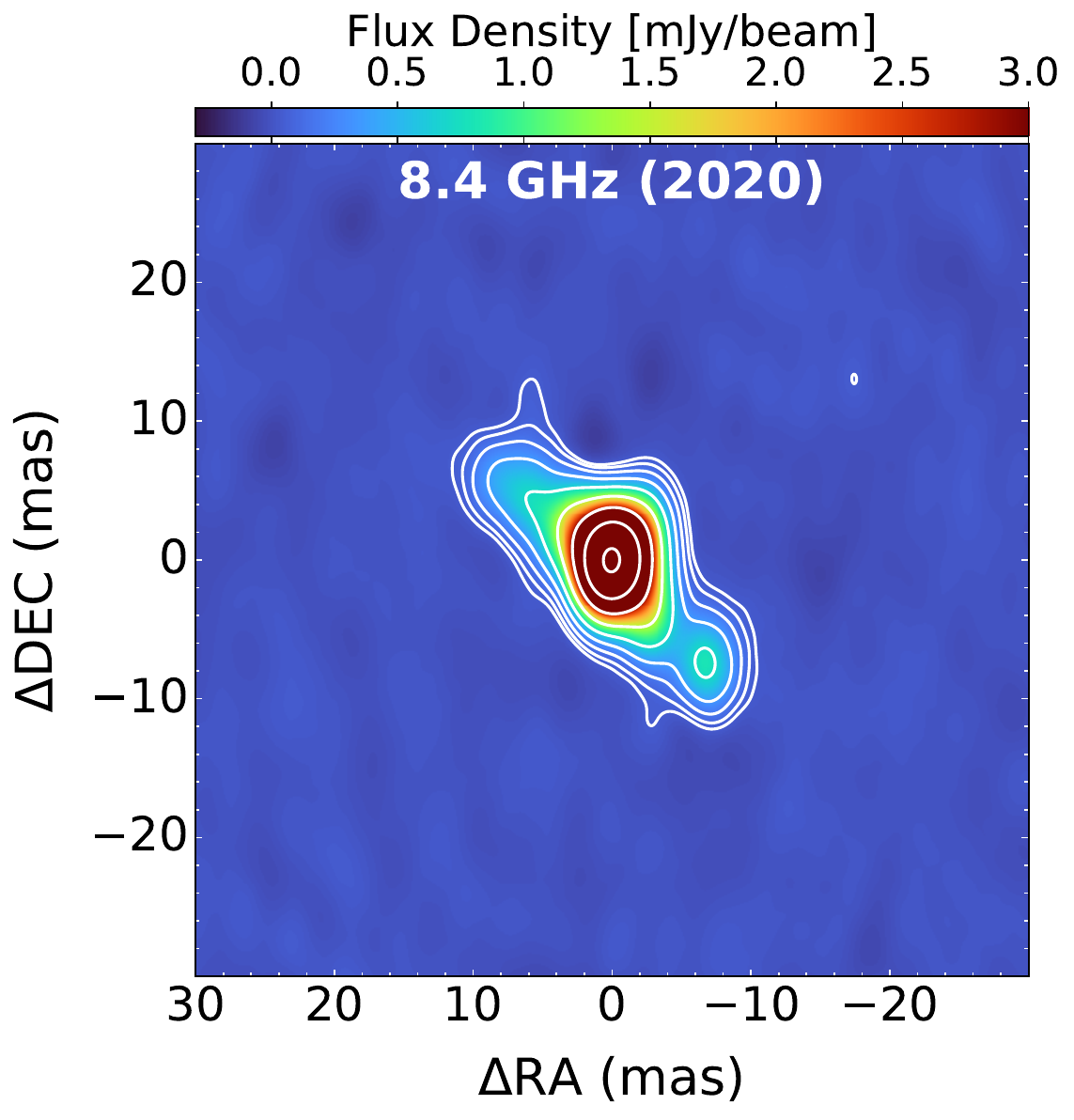}
    \includegraphics[width=0.32\linewidth]{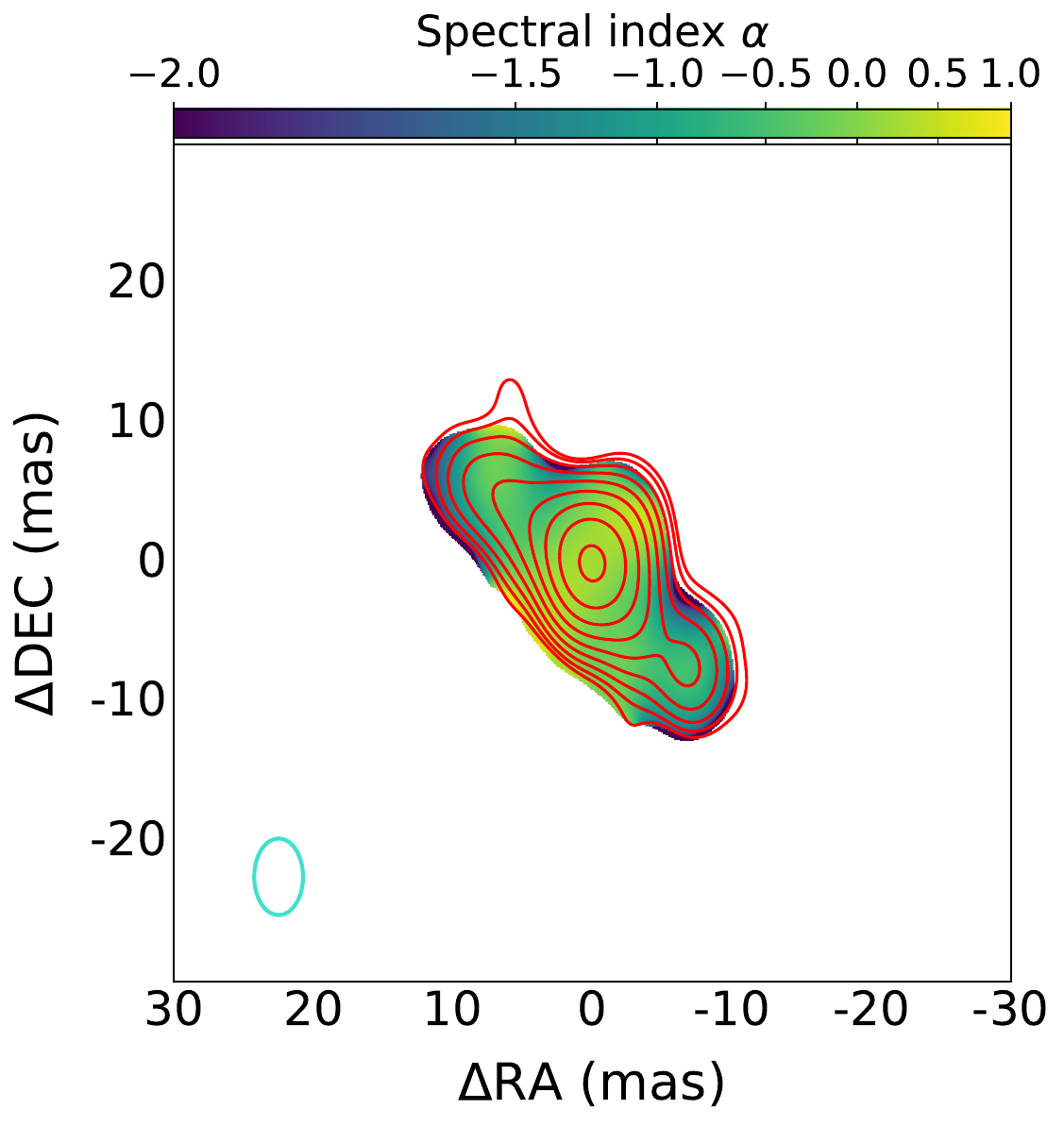}\\
    \includegraphics[width=0.32\linewidth]{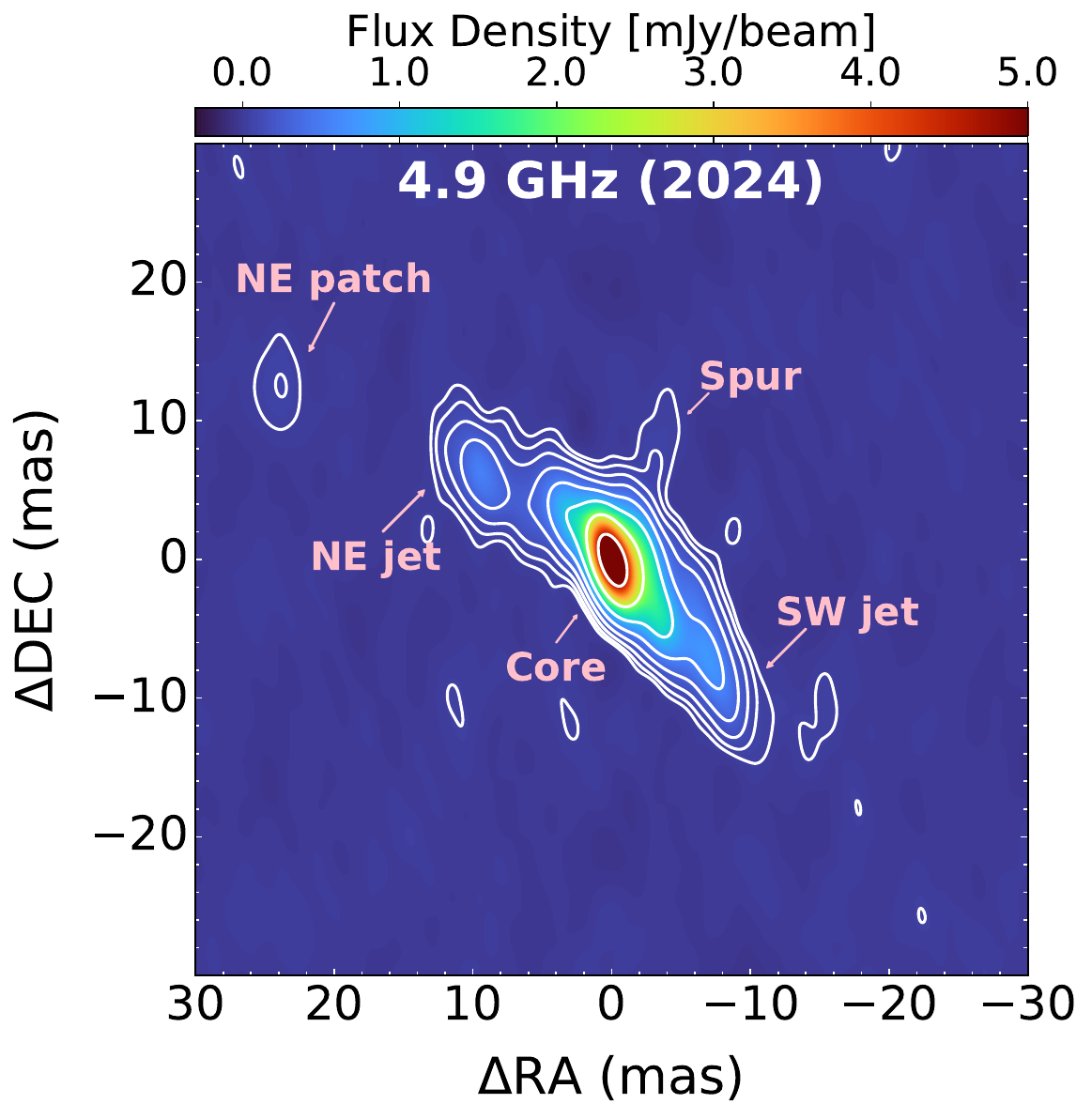}
    \includegraphics[width=0.32\linewidth]{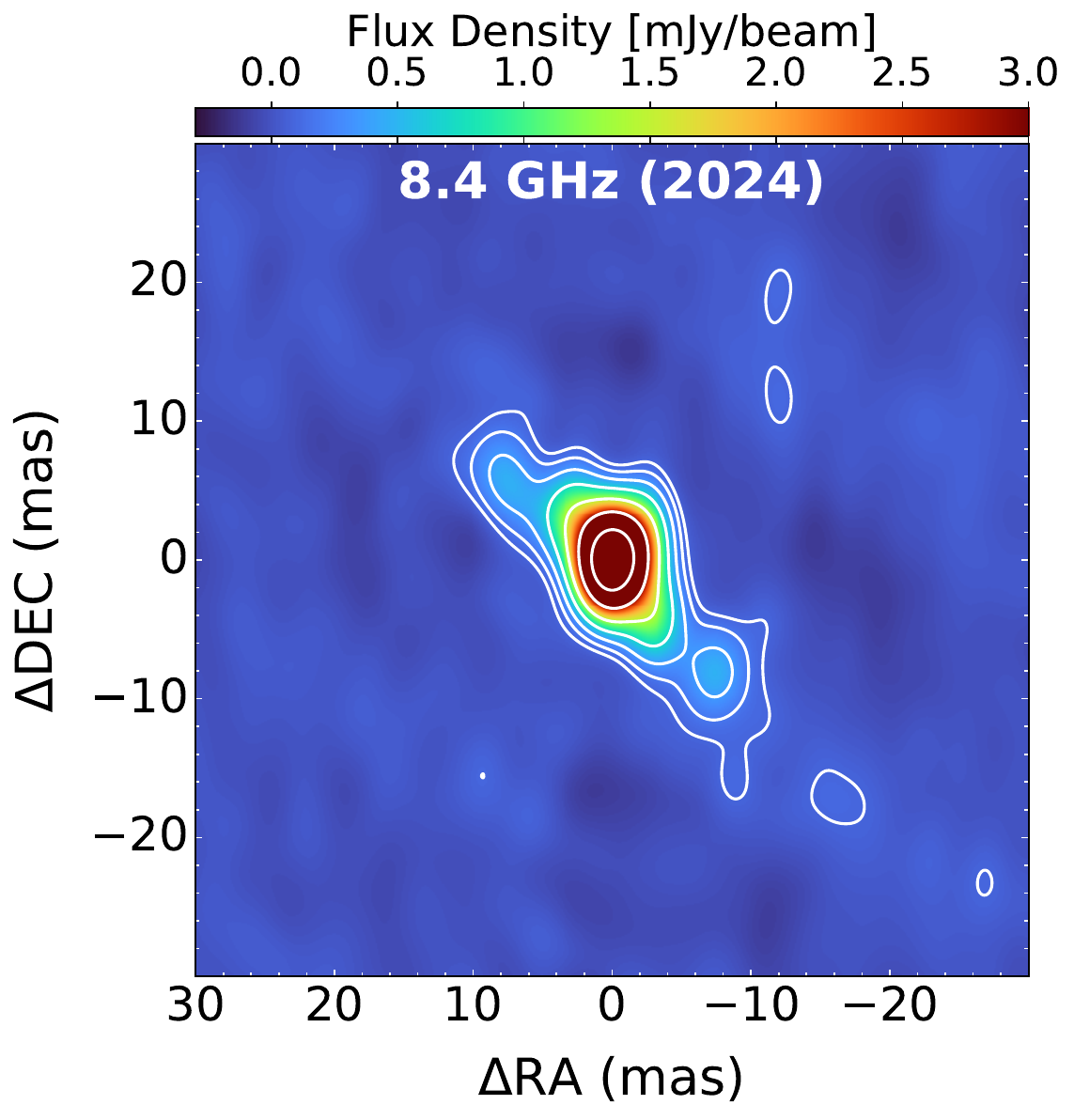}
    \includegraphics[width=0.32\linewidth]{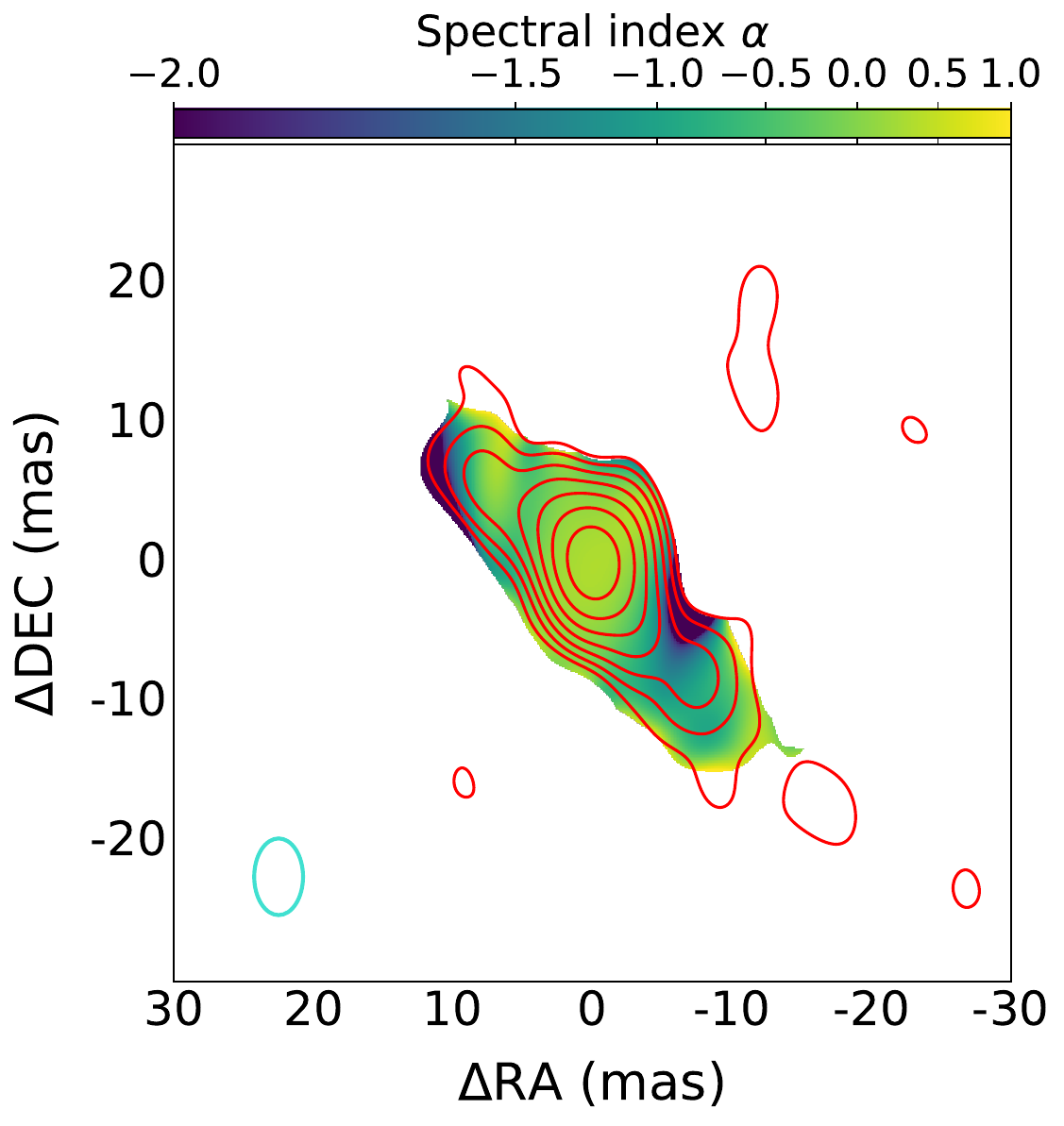}
    \caption{Multi-epoch VLBA images of NGC~5044. {\it Left} \& {\it Middle panels}: we show the 2020 (top) and 2024 (bottom) 4.9 GHz (C band; left) and 8.4 GHz (X band; middle) images of NGC~5044, at matching angular resolution between the two epochs at each frequency. The panels have matching field of view (60$\times$60~mas$^{2}$ box, or 9$\times$9 pc$^{2}$), and matching color-scale at each frequency. The angular resolution is about 5$\times$2 mas (see Table~\ref{tab:data} for the exact values and for the rms noise levels). White contours are drawn at 3$\sigma_{rms}$ and increase by a factor of 2. Labels in the bottom left panel highlight structures discussed in the text. {\it Right panels:} we show the 4.9 GHz - 8.4 GHz spectral index maps from 2020 (top) and 2024 (bottom) data (see Section~\ref{subsub:continuumCX} for details). The beam FWHM (cyan ellipse) is 5.5~mas$\times$3.5~mas, at P.A. of 0$^{\circ}$. The red contours are drawn at 3$\sigma_{rms}$ of the X band image used to produce the spectral index map (very similar to the middle panel), and increase by a factor of two.}
    \label{fig:CXepochs}
\end{figure*}

\begin{figure}[ht!]
    \centering
        \includegraphics[width=\linewidth]{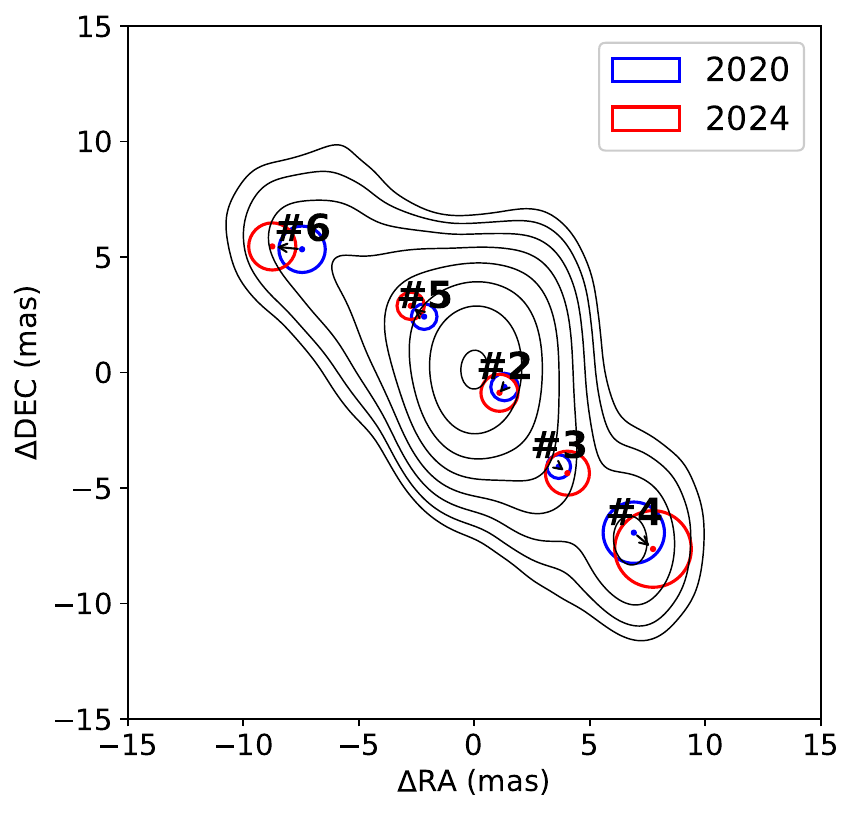}
    \caption{Jet proper motions in NGC 5044. Results for \texttt{Difmap}'s \texttt{modelfit} task applied to the 4.9 GHz and 8.4 GHz data from 2020 and 2024. Jet components from the two epochs are shown as blue (2020) or red (2024) dots, with surrounding circles showing the best-fit extent of the circular Gaussian used to fit the data. We stress that the uncertainty on the position of the components (a few 10s of $\mu$as) is smaller than the circles. The black contours are from the 8.4 GHz data from 2020.}\label{fig:CXdiff}
\end{figure}
\subsection{C and X band observations}
\subsubsection{Continuum analysis}\label{subsub:continuumCX}
We show in Figure~\ref{fig:CXepochs} the 2020 (top) and 2024 (bottom) 4.9 GHz and 8.4 GHz images of NGC~5044, at matching angular resolution between the two epochs at each frequency (4.8$\times$1.9~mas at 4.9 GHz, and 5.0$\times$2.9~mas at 8.4 GHz). The overall recovered structure is very similar across the images: a bright core with symmetric northeast-southwest jets. As visible in Figure~\ref{fig:CXepochs}, the 4.9 GHz VLBA observations from 2020 and 2024 recover a “spur" of radio emission northwest of the radio core at $3\sigma_{rms}, 6\sigma_{rms}$, best visible in the 2024 epoch (which is 40\% more sensitive). Interestingly, the extent of this spur of radio emission is aligned with the kpc-scale northern cavities and the excess extended radio emission perpendicular to the jets in the 1.4 GHz continuum image of Figure~\ref{fig:continuum}. We also find another patch of the northeast jet at 4.9 GHz, at $\Delta$RA$=$23~mas, $\Delta$DEC$=$12~mas from the core, that was already present in the 2020 data and is confirmed in the 2024 data at $6\sigma_{rms}$ confidence. The existence of this patch at $\sim$26~mas (4~pc) from the core is consistent with the larger extent of the jets at 1.4 GHz discussed in Section~\ref{subsec:1.4continuum} and illustrated in Figure~\ref{fig:continuum} and Figure~\ref{fig:LLS-jets}. \\ 
\par We note that the flux density of the core varied by approximately 20\% at both 4.9 GHz and 8.4 GHz, as visible in the images shown in Figure~\ref{fig:CXepochs}. This factor is larger than the systematic flux density uncertainty of the observations ($\sim$5\%), and suggests that the core of NGC~5044 experiences variability in radio over timescales of a few years or less (the time baseline of our observations is $\Delta t = 4.5$~yr). This is unsurprising for the radio cores of cluster or group central galaxies \citep{hogan2015b,rose2022}. It is also interesting that clear flux variability of $\sim20\%$ amplitude on months timescales in the sub-millimeter band has been reported recently in \citet{schellenberger2024}. \\

\par The total VLBA flux densities of the source at the different epochs are $S_{4.9}^{2020} = 21.7\pm1.9$~mJy, $S_{8.4}^{2020} = 20.7\pm1.7$~mJy, $S_{4.9}^{2024} = 17.4\pm1.1$~mJy, and $S_{8.4}^{2024} = 16.2\pm0.9$~mJy. These measurements give integrated spectral indices of $\alpha\sim 0.1$. Previously, \citet{schellenberger2021} reported a spectral index map between the 4.9 GHz and 8.4 GHz data from 2020, showing an overall flat spectral index with inverted (positive) values close to the core and flat or slightly steep (negative) values in the region of the jets. Using observations from both epochs, we produced the pixel-wise spectral maps by matching the (\emph{u,v})-coverage of the 4.9 GHz and 8.4 GHz data and imaging with a common angular resolution of 5.5~mas~$\times$~3.5~mas. The spectral index map was derived after imposing 3$\sigma_{rms}$ cuts in surface brightness at both frequencies. The resulting spectral index images are shown in Figure~\ref{fig:CXepochs} (right panels) overlaid with 8.4 GHz contours. The spectral index image from the first epoch is identical to the one shown in \citet{schellenberger2021}. The spectral index image from the second epoch is consistent with that from 2020, showing inverted spectrum, $\alpha>0$ values at the positions of the core and steeper spectrum, $\alpha \leq 0$ values in correspondence to the jets. Overall, placing a beam-sized region on top of the core, we measure spectral indices of $\alpha^{2020}_{core} = 0.20\pm0.07$ and $\alpha^{2024}_{core} = 0.23\pm0.04$. In the combined region of the jets, we find $\alpha^{2020}_{jets} = -0.79\pm0.35$ and $\alpha^{2024}_{jets} = -0.81\pm0.30$. This is consistent with the typical properties of pc-scale synchrotron emission of AGN -- for comparison, the analysis of 190 MOJAVE AGN \citep{hovatta2014} reported average core and jet spectral index of $+0.22$ and $-1.04$, respectively.

\begin{center}
\begin{table*}[t]
     \centering
     \caption{Results from \texttt{Difmap} fitting of the combined 4.9 GHz and 8.4 GHz visibilities from 2020 and 2024 (Figure~\ref{fig:CXdiff} and Section~\ref{subsub:propermotion}).}\label{tab:difmap}\renewcommand*{\arraystretch}{1.5}
     \begin{tabular}{lccccccccccc}

     \hline     
    & Comp. & Epoch & $S_{6.7}$ & SNR & $\Delta$RA & $\Delta$DEC & $d$ & $r_{proj,c}$ & $\Delta$r & $\mu$& $\beta = v/c$ \\
   & &  & mJy & & mas & mas & pc & pc & mas & mas/yr &  \\

\hline

  \multirow{6}{*}{SW jet} &  \multirow{2}{*}{\#2} & 2020 & 7.71 & 385& $1.3$ & $-0.6$& 0.18 & 0.24 & \multirow{2}{*}{0.42$\pm$0.03}& \multirow{2}{*}{0.09$\pm$0.01}& \multirow{2}{*}{0.045$\pm$0.003}\\
\cline{3-9}
  & & 2024 & 6.53 & 326  & $1.1$ & $-0.9$ & 0.24& 0.24 & & & \\

\cline{2-12}

  &  \multirow{2}{*}{\#3} & 2020 & 1.74  & 87&  $3.7$ & $-4.1$ & 0.15& 0.86 & \multirow{2}{*}{0.47$\pm$0.11}& \multirow{2}{*}{0.10$\pm$0.02}& \multirow{2}{*}{0.051$\pm$0.012}\\
\cline{3-9}
  & & 2024 & 2.15 & 107  & $4.0$ & $-4.4$ & 0.29& 0.92 & & &\\

\cline{2-12}

  &  \multirow{2}{*}{\#4} & 2020 & 0.89 & 44 &  $6.9$ & $-6.9$& 0.40& 1.51 & \multirow{2}{*}{1.09$\pm$0.19}& \multirow{2}{*}{0.24$\pm$0.04}& \multirow{2}{*}{0.118$\pm$0.021}\\
\cline{3-9}
  & & 2024 & 1.60 & 80  & $7.7$ & $-7.6$& 0.50& 1.66 & & &\\

\hline

  \multirow{6}{*}{NE jet} &  \multirow{2}{*}{\#5} & 2020 & 2.36 & 118&  $-2.2$ & $2.4$ & 0.17& 0.45 & \multirow{2}{*}{0.76$\pm$0.11}& \multirow{2}{*}{0.17$\pm$0.03}& \multirow{2}{*}{0.083$\pm$0.012}\\
\cline{3-9}
  & & 2024 & 1.57  &79 & $-2.7$ & $2.9$ & 0.18& 0.57 & & &\\

\cline{2-12}

  &  \multirow{2}{*}{\#6} & 2020 & 1.60 & 80&  $-7.5$ & $5.3$& 0.30& 1.34 & \multirow{2}{*}{1.35$\pm$0.18}& \multirow{2}{*}{0.30$\pm$0.04}& \multirow{2}{*}{0.147$\pm$0.019}\\
\cline{3-9}
  & & 2024 & 1.00  &50 & $-8.7$ & $5.5$ & 0.31& 1.52 & & &\\

\cline{2-12}

&  \multirow{2}{*}{\#7$^{(*)}$} & 2020 & 0.41 & 20 & $-25.6$ & $14.8$ & 0.87& 4.40& \multirow{2}{*}{2.05$\pm$1.05}& \multirow{2}{*}{0.46$\pm$0.19}& \multirow{2}{*}{0.223$\pm$0.109}\\
\cline{3-9}
  & & 2024 & 0.20 & 10 & $-27.7$ & $14.4$ & 0.93& 4.67 & & &\\

\hline
  
\end{tabular}
 \vspace{4pt}
\tablecomments{$^{(*)}$ NE patch, detected at C band only. Columns: (1) Structure; (2) label of the component; (3) epoch; (4) flux density at the intermediate 6.7~GHz frequency; (5) signal-to-noise ratio of the component; (6) offset in right ascension from the core at (0,0); (7) offset in declination from the core at (0,0); (8) extent (FWHM) of the circular Gaussian used to fit the component; (9) projected distance of the component from the core at (0,0), measured from the epoch 2024; (10) distance traveled by the component between the two epochs; (11) proper motion in angular units; (12) advance speed in units of the speed of light. }
 \end{table*}
\end{center}

\subsubsection{Proper motion analysis}\label{subsub:propermotion}
With the availability of two epochs, we can study the proper motion of the jets in NGC~5044. The most reliable method to measure proper motions is to robustly identify the position of the source sub-components at each epoch/frequency, and then measure the positional offsets as differences in RA and DEC relative to a common center \citep{mirabel1994}. An established approach is to use the software \texttt{Difmap} \citep{difmap1997} to identify the components in the (\emph{u,v})-plane, by directly fitting the interferometric visibilities and without relying on image reconstruction \citep[e.g.,][]{giroletti2003,giroletti2005}. Adopting this approach, we combined the 4.9 GHz and 8.4 GHz visibilities at each epoch to maximize the SNR, and fitted the visibilities in \texttt{Difmap} with sets of circular Gaussians. 
We found that the jet structure is well described in the (\emph{u,v})-plane by a model with three components in the southwest jet, and two components in the northwest jet. The position and size of the components are illustrated in Figure~\ref{fig:CXdiff}, while the relevant parameters of the fit for the two epochs are reported in Table~\ref{tab:difmap} (flux densities are reported at the intermediate frequency of 6.7~GHz). 
We also report results from a fit to the 4.9 GHz data only, where we can also recover the extended NE patch visible in Figure~\ref{fig:CXepochs} (reported as component \#7 in Table~\ref{tab:difmap}). Statistical errors on the parameters obtained from the \texttt{Difmap} fit may underestimate the real uncertainties \citep[e.g.,][]{deller2018}, therefore, we computed positional and size uncertainties as the beam FWHM divided by the SNR of the detection. 

Adopting the position of the core at (0,0) as reference, we compared the position of the jet components and derived the relative distance over the 4.5 years and the velocity in mas/yr and in units of $c$ (see Table~\ref{tab:difmap}). We find that velocities are relativistic but sub-luminal. This is consistent with the source lying close to the plane of the sky (see also the analysis of \citealt{schellenberger2021,ubertosi2024a}); if significant inclination was present, super-luminal speeds could be observed. By averaging together the relative velocities, we obtain an average source expansion speed of $\langle v_{j} \rangle = (0.10\pm0.02)\,c$. 

An independent constraint on the jet speed can be obtained from the jet-to-counterjet flux density ratios, by assuming a symmetric ejection of jet knots, the observation of which is affected by Doppler boosting. For a pair of knots with flux density ratio R (see e.g., \citealt{scheuer_1979}):
\begin{equation}\label{eq:rfromtheta}
R = \left(\frac{1 + \beta \cos\theta}{1 - \beta \cos\theta}\right)^{k-\alpha},
\end{equation}
where $\beta = v/c$ is the bulk velocity of the emitting plasma, $\theta$ is the inclination angle relative to the line of sight, $\alpha$ is the spectral index, and $k=3$ ($k=2$) for discrete jet knots (continuous jets). From the above equation, the jet speed can be expressed as:
\begin{equation}
\label{eq:vfromR}
v = \left(\frac{R^{1/(k-\alpha)} - 1}{R^{1/(k-\alpha)} + 1} \times \frac{1}{\cos\theta}\right) c\,\,.
\end{equation}
Adopting $k=3$, and a spectral index $\alpha=-0.8$ as measured from the data (Section \ref{subsub:continuumCX}), we estimated the bulk velocity of the opposite knot pairs (\#3 \& \#5) and (\#4 \& \#6) using Equation \ref{eq:vfromR} (see Table \ref{tab:difmap}). Given the degeneracy between the jet velocity $\beta$ and the inclination angle $\theta$ in the jet-to-counterjet ratio, and the lack of independent constraints on the source orientation\footnote{For comparison, we note that \citet{ubertosi2024a} estimated $\theta=4.4^{\circ}\pm5.3^{\circ}$ from Equation \ref{eq:rfromtheta} by assuming $0.6\leq\beta\leq1$ (higher than our proper motion-based measurement).}, we adopted a representative value of $\theta = 45^{\circ}$. For each knot pair, the calculations were done separately for the two epochs and then averaged together, using the dispersion around the mean as an estimate of the uncertainty. We derived velocities of $v = (0.06 \pm 0.01)\,c$ for knot pair~1 (\#3 \& \#5) and $v = (0.10 \pm 0.02)\,c$ for knot pair~2 (\#4 \& \#6). These values are in agreement with the apparent speeds inferred from the proper motion analysis above (average expansion velocity of $\langle v_j \rangle = (0.10 \pm 0.02)\,c$), thus supporting the idea the jets of NGC~5044 are oriented close to the plane of the sky (see also \citealt{schellenberger2021,ubertosi2024a}) and are mildly relativistic.

We note that, with only two epochs, it is easy to over-interpret the finer details of each component’s motion. For instance, the apparent lack of outward motion in component \#2, that seems to be moving predominantly tangentially rather than radially with respect to the jet axis, may suggest complex motions e.g.,  precession of the jet axis. While this possibility is worth mentioning, confirmation of such trends would require long-term monitoring of NGC~5044 with additional epochs.

\begin{figure*}[ht!]
    \centering
    \includegraphics[width=\linewidth]{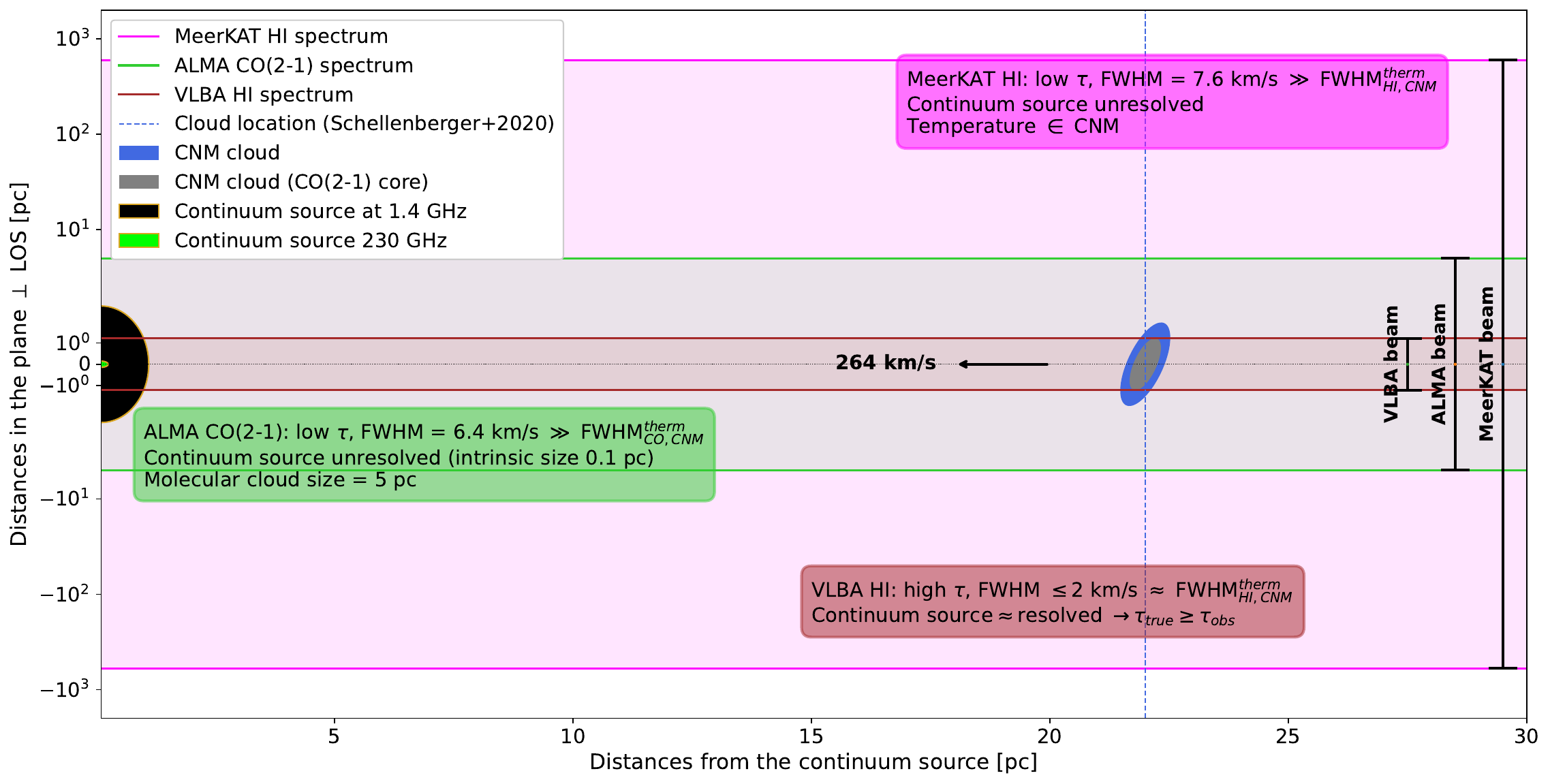}
    \caption{This simplified sketch shows the scenario proposed to explain the properties of the atomic and molecular gas absorption line at 264~km~s$^{-1}$ in the spectrum against the radio core of NGC 5044. Distances along the line of sight (LOS) are shown on the horizontal axis and distances perpendicular to the LOS on the vertical axis (both in pc). The continuum source is drawn at the origin; the different sizes of the continuum source at 1.4\,GHz (black ellipse) and 230\,GHz (green ellipse) reflect the different observed sizes of the continuum source in VLBA (this work) and ALMA \citep{schellenberger2021} data. A cold neutral medium (CNM) cloud at $\sim$22\,pc from the continuum source (blue ellipse; dashed line) contains a denser molecular core traced by CO(2-1) absorption (grey ellipse). The arrow indicates the $\sim$264\,km\,s$^{-1}$ infall velocity of the absorbing cloud. Shaded bands illustrate the transverse scales sampled by the MeerKAT H\,\textsc{i} (magenta), ALMA CO(2--1) (green), and VLBA H\,\textsc{i} (brown) angular resolutions (restoring beams). MeerKAT and ALMA detect absorption lines against an unresolved continuum emission with low observed peak optical depths and broad FWHM ($\gg$ thermal CNM widths), while the VLBA partially resolves the continuum source, revealing higher peak optical depth and a narrow H I line (FWHM $\lesssim$2\,km\,s$^{-1}$).}
    \label{fig:absorption_model}
\end{figure*}

\section{DISCUSSION}\label{sec:discussion}
\subsection{Cold-mode AGN feeding at pc scales}\label{subsec:agnfeeding}
Detections of molecular or atomic narrow lines with redshifts of a few $\sim$100 km~s$^{-1}$ have been previously reported in a number of cluster-central radio galaxies (e.g., \citealt{tremblay2016,rose2023}). These lines likely trace clouds that are infalling and eventually feeding the central AGN (e.g., \citealt{rose2023}; see also \citealt{schellenberger2020} for the case of NGC~5044). Another argument routinely made in line absorption studies is that these clouds reside close to the AGN. In NGC~5044, our VLBA data reveal narrow atomic HI lines against the core (redshifted at 264~km~s$^{-1}$ and 232~km~s$^{-1}$). In the following, we restrict our discussion to the line at 264~km~s$^{-1}$ against the core, which, as reported in Section \ref{subsec:spectrallineanalysis}, is detected at high significance. 
\\\par Similarly to the ALMA and MeerKAT detections of CO and HI, respectively, in this galaxy, the VLBA detection likely traces individual HI clouds lying along our line of sight to the radio core. The close agreement in the 264 km~s$^{-1}$ line velocity between the VLBA, MeerKAT, and ALMA data suggests that we are observing the same gas cloud in different phases, spanning a range of densities and temperatures from atomic to molecular. However, there are some key differences in the properties of the gas and of the continuum source, that we list here:
\begin{itemize}[leftmargin=*,noitemsep]
    \item {\it The continuum source:} spatially unresolved in MeerKAT and ALMA data, but clearly resolved in the VLBA data. The continuum source size also changes with frequency. At 1.4~GHz, the continuum source has a size of $\approx$10~pc, and is thus unresolved by MeerKAT (beam FWHM $\sim$750~pc) but resolved by the VLBA (beam FWHM $\sim$2.4~pc). If the continuum source at 230~GHz had the same extent as at 1.4~GHz, the ALMA beam (FWHM $\sim$7.5~pc) would have nearly resolved it. Indeed, \citet{schellenberger2024} reported a size of $\sim$1.5~pc for the continuum source at 230~GHz. Therefore, the continuum source size decreases with increasing frequency, which is compatible with the difference in size between the mm-emitting section of the accretion flow \citep{schellenberger2024} and the GHz emitting region of the accretion flow plus GHz emission from the radio jets. 
    \item {\it FWHM of the narrow lines:} All detected lines are narrow - close to 1~km~s$^{-1}$ (VLBA) or a few km~s$^{-1}$ (ALMA and MeerKAT). These widths should be compared with the thermal broadening expected for the relevant atomic and molecular transitions. In the cold neutral medium (CNM; $T \sim 100$~K), thermal broadening alone would produce a CO line with FWHM $\approx$0.3--0.4~km~s$^{-1}$, compared to the $\sim$7~km~s$^{-1}$ width measured by \citet{schellenberger2020}. Thus, the velocity dispersion in the cloud is largely due to turbulence and internal motion. The HI line detected by MeerKAT confirms that the gas is in the CNM \citep{rajpurohit2025}; for $T \le 100$~K, the expected thermal FWHM of HI is 1.5--2~km~s$^{-1}$. The line broadening seen by MeerKAT is therefore also attributable to turbulent motions. In contrast, the VLBA HI line is as narrow as the predicted thermal width (providing an upper limit of $T \sim 60$~K for the gas). 
    \item {\it Optical thickness of the cloud:} The HI lines detected by MeerKAT have relatively low peak optical depth values ($\tau_{HI,\,M}\sim0.03$; optically thin regime), as do the CO(2–1) lines detected by ALMA ($\tau_{CO,\,A}\sim0.2$). The VLBA detections, however, are closer to the regime of optically thick gas ($\tau_{HI,\,V} \approx 1$). The true optical depth may be even higher once the covering factor of the clouds is taken into account: for a resolved continuum source, adopting $0.5 \le c_f \le 0.75$ would imply $\tau_{HI,\,V} \gg 1$ for the HI line. This is consistent with the higher angular resolution of the VLBA relative to MeerKAT. The fact that this is not observed with ALMA is due to the smaller size of the continuum source at 230~GHz.
\end{itemize}

We thus propose the following scenario (Figure~\ref{fig:absorption_model}): a set of clouds is located within the sphere of influence of the SMBH ($r \lessapprox 25$~pc; \citealt{schellenberger2020}). The cloud with infall velocity of 264~km~s$^{-1}$ consists of a molecular core surrounded by atomic layers, with a size of $\sim$5~pc \citep{schellenberger2020,rajpurohit2025} and a temperature typical of the CNM (10--100~K). The cloud produces an absorption line against the continuum source, whose size is $\sim$10~pc at 1.4~GHz (HI transition) and $\sim$1.5~pc at 230~GHz (CO(2–1) transition). 
The extremely narrow absorption line detected by the VLBA -- whose FWHM is consistent with thermal broadening alone -- is explained by the high optical depth of the gas: at low temperatures, optically thick gas produces intrinsically narrow absorption lines when the absorption arises only from a thin outer layer \citep{koribalski1996}, as is likely the case here. Conversely, the broader ALMA and MeerKAT line is consistent with turbulent broadening in optically thin gas.

We note here that the unresolved core continuum flux density at 1.4\,GHz is $S_{1.4} \simeq 8.4$\,mJy (or $\sim$7\,mJy in the \texttt{R$=-2$, 10-45~M$\lambda$} image, see Figure~\ref{fig:LLS-jets}). This emission represents the combined contribution of synchrotron from the inner jet knots, together with low-frequency emission from the advection-dominated accretion flow (ADAF) investigated by \citet{schellenberger2020,schellenberger2024} for NGC~5044. Interestingly, extrapolating the best-fit ADAF spectral energy model from \citet{schellenberger2020,schellenberger2024} to 1.4~GHz yields an expected ADAF flux density of $\approx$8\,mJy at 1.4\,GHz, in agreement with the observed flux density.

It is also interesting to compare the estimates of the mass and size of the atomic and molecular gas clouds. \citet{schellenberger2020} used the velocity dispersion $\sigma_{v}$ of the 258~km~s$^{-1}$ and the 264~km~s$^{-1}$ absorption lines against the core to infer the size $d$ of the clouds, using the $\sigma_{v}-d$ scaling relation for molecular gas clouds from \citet{solomon1987}. Based on this relation, they determined sizes of $d \sim 5$~pc for the molecular gas clouds. Then, assuming that the absorption clouds of radius $R = d/2$ are in virial equilibrium:
\begin{equation}
    \label{eq:virialeq}
    M_{cloud} \approx \frac{R\sigma_{v}^{2}}{G}\,,
\end{equation}
\citet{schellenberger2020} determined a mass for each molecular clouds of $M_{mol}^{cloud}\sim7\times10^{3}$~M$_{\odot}$. 
\\ \indent We can provide similar estimates for the atomic gas line at 264~km~s$^{-1}$ seen in MeerKAT and VLBA data. \citet{wolfire2003} and \citet{kalberla2009} defined a $\sigma_{v}-d$ relation for atomic clouds:
\begin{equation}\label{eq:sizeHI}
    d = \left(\frac{\sigma_{v}}{1.3\,\text{km s}^{-1}}\right)^{3}\quad\text{[pc]}
\end{equation}
Focusing on the 264~km~s$^{-1}$ line in the MeerKAT spectrum, for which $\sigma_{v} = 3$~km~s$^{-1}$, Equation~\ref{eq:sizeHI} returns $d = 12$~pc. As the VLBA line arises from the outermost layer of optically thick atomic gas, its velocity dispersion cannot be related to a characteristic size (for completeness, with $\sigma_{v} = 0.7$~km~s$^{-1}$, we would obtain a size of $d=0.2$~pc). 
Assuming that the absorption cloud seen by MeerKAT is in virial equilibrium, its mass can be computed from Equation~\ref{eq:virialeq}. We find $M_{cloud}^{HI} = 1.2\times10^{4}$~M$_{\odot}$ (line A), close to the ALMA mass. 

While we cannot directly measure the number density of the absorbing atomic gas, we can constrain it by combining the HI column density limits with the characteristic cloud sizes inferred above (see e.g., \citealt{wolfire2003}). From the analysis of the 264 km~s$^{-1}$ HI line (Section~\ref{subsec:spectrallineanalysis}), the HI column density is constrained to be $N_{\text{HI,abs}} \lesssim 2\times10^{21}$~cm$^{-2}$, which implies an upper limit on the average volume density of $n_{\mathrm{HI}} \lesssim N(\mathrm{HI})/d \sim 6\times10^{2}/[d(\mathrm{pc})]$~cm$^{-3}$. For cloud sizes of $d \sim 12$~pc, this would yield $n_{\mathrm{HI}} \lesssim 50$~cm$^{-3}$. An independent estimate can be obtained from the cloud mass, which gives an average density of $n \sim M_{cloud}^{HI}/V \sim 2.4\times10^{5}/[d(\mathrm{pc})]^{3}$~cm$^{-3}$, corresponding to $n \sim 140$~cm$^{-3}$ for $d \sim 12$~pc. Given the (highly uncertain) assumptions on the cloud size, the virial equilibrium, and the filling factor, these density estimates can be considered as broadly consistent.

Our results, informed by the previous analysis of \citet{schellenberger2020, rajpurohit2025}, demonstrate that the gas within a few tens of pc of the SMBH in NGC~5044 is clearly multiphase, with atomic and molecular components coexisting in individual cold (10–100~K) clouds of a few pc sizes. These clouds are likely infalling toward the AGN, potentially feeding it, and their presence at such small radii highlights the importance of compact, dense multiphase gas structures in the fueling of cluster- and group-central radio galaxies.

\begin{figure*}[ht!]
    \centering
    \includegraphics[width=\linewidth]{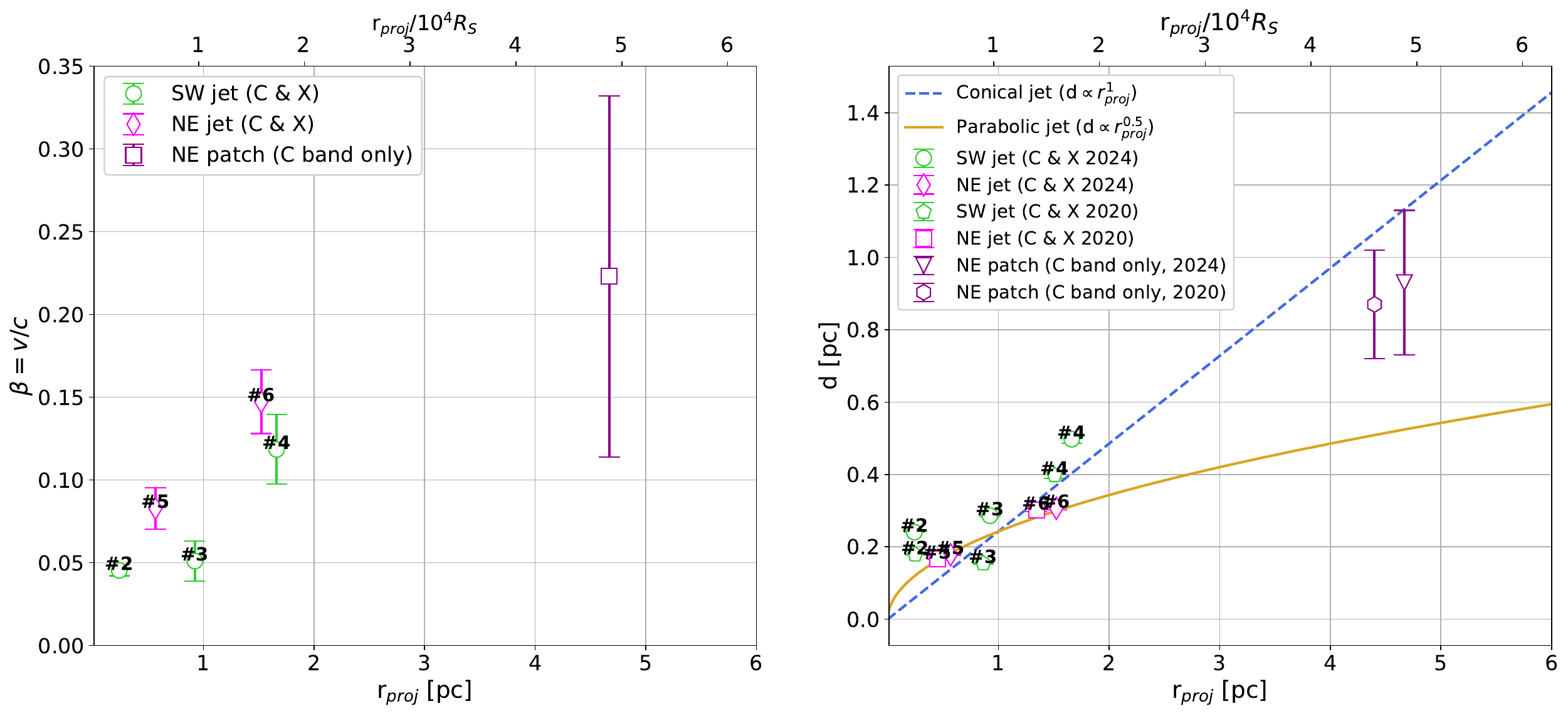}
    \caption{Velocity and jet width radial profiles of the jets in NGC~5044. {\it Left:} The advance speed of the jet components listed in Table~\ref{tab:difmap} is plotted as a function of their projected distance from the core. We also include the constraint (although with large errorbars) given by the NE patch. {\it Right:} The width of the jet components listed in Table~\ref{tab:difmap} is plotted as a function of their projected distance from the core. Overlaid are theoretical collimation profiles for a perfectly conical jet (dashed blue line) and for a perfectly parabolic jet (solid gold line). In both panels, the top x-axis shows the distance from the core in units of $10^{4} R_{s}$ (Schwarzschild radius), computed assuming a SMBH mass of $10^{9}$~M$_{\odot}$.}
    \label{fig:jetcollimation}
\end{figure*}

\subsection{Jet collimation and energetics}
The clear proper motions of the jet components in NGC~5044 suggest that the AGN is not in a fading phase, but rather active and propelling the jets at sub-relativistic speeds. Based on the maximum size of the NE-SW jets ($L_{j}\sim5.5$~pc, based on the L band images, see Figure~\ref{fig:continuum} and Figure~\ref{fig:LLS-jets}), and the average expansion speed of the source measured from proper motions ($\langle v_{j}\rangle\sim0.1\,c$, see Section \ref{subsub:propermotion}), we can estimate a dynamical age:
\begin{equation}\label{eq:tdyn}
    t_{dyn} = \frac{L_{j}}{\langle v_{j}\rangle} \approx 180\,\,\text{yr}.
\end{equation}
This is a rough estimate, as the velocity of expansion is not constant in the radial range that we considered ($\leq$6~pc), as suggested by our results from Section~\ref{subsub:propermotion}. 
\\\par Thanks to the availability of proper motion measurements at different distances from the center in NGC~5044, we can build radial profiles of advance speed ($\beta(r_{proj}) = v(r_{proj})/c$) and jet width ($d(r_{proj})$). These have been used to constrain the 3D shape of the jets and the collimation processes occurring close to the SMBH (e.g., \citealt{park2019,kovalev2020,boccardi2021_proceeding,boccardi2021}). The resulting Figure~\ref{fig:jetcollimation} shows the $\beta(r_{proj})$ and $d(r_{proj})$ radial profiles in the left and right panels, respectively, based on the components \#2 -- \#7 in Table~\ref{tab:difmap}. 
\\\indent Although the availability of only a few radial bins prevents us from drawing definitive conclusions, we tentatively observe that the velocity profile suggests an outward acceleration of the jet flow, from $\sim$0.05~c at 0.5~pc from the core to $\sim$0.15~c at 1.6~pc from the core. At larger distances, the NE patch has a somewhat larger advance speed ($\sim0.22\,c$), however the large uncertainty makes it difficult to assess changes in the slope of $\beta(r_{proj})$. The jet width profile is more informative. Below $r_{proj}\leq1.5$~pc, the jet width shows a mild increase ($\leq$0.1 pc width per pc distance). Beyond $r_{proj}\approx1.5$~pc, the jet width increases more rapidly at a rate of $\approx0.3$pc per pc. Changes in the slope of the jet width profiles have been associated with changes in the collimation and acceleration of the jets for several classes of AGN \citep{boccardi2019,boccardi2021}. For radiatively inefficient AGN such as NGC~5044 \citep{schellenberger2024}, a transition in the profile is typically observed as follows:
\begin{equation}
d(r_{proj}) \propto \left\{
\begin{array}{ll}
r_{proj}^{0.5} & \text{(parabolic) if } r_{proj}\lessapprox10^{5} R_{S},\\
r_{proj}^{1}, & \text{(conical) if } r_{proj}\gtrapprox10^{5} R_{S}
\end{array}
\right.
\end{equation}
with $R_{S}$ being the Schwarzschild radius of the SMBH: 
\begin{equation}
    R_{S} = \frac{2GM_{\rm BH}}{c^{2}}
\end{equation}
As pointed out in Section \ref{sec:intro}, various estimates of the SMBH mass for NGC~5044 exist (see e.g., \citealt{david2009,diniz2017,schellenberger2024}). Here we adopt an average SMBH mass of $\approx10^{9}$~M$_{\odot}$ to compute the Schwarzschild radius of the SMBH. This choice does not strongly influence the results discussed below, which are valid for order-of-magnitude estimates of $r/R_{S}$. We show the radial distance as a function of $R_{S}$ as the top x-axis of Figure~\ref{fig:jetcollimation}. We also overlay tracks for conical and parabolic jet collimation profiles in the right panel. The flatter and steeper $d(r_{proj})$ profile of NGC~5044 below and beyond $r_{proj}\approx1.5$~pc is suggestive of a transition from a close to parabolic to a conical collimation profile a few $\approx10^{4}\,R_{S}$ from the core, as found for other radiatively inefficient AGN. In terms of velocity radial profile, one would expect more acceleration to occur in the parabolic regime (steeper $\beta(r_{proj})$), followed by a progressive flattening of $\beta(r_{proj})$ in the conical regime (e.g., \citealt{park2019}). The single point from the NE patch in the left panel of Figure~\ref{fig:jetcollimation} might suggest that some deceleration is present beyond $r_{proj}\approx1.5$~pc, but additional constraints from more sensitive, multi-epoch observations are necessary to confirm these indications.
\\\par With the available data and information, we can also provide a rough estimate of the energetics of the compact radio source. These estimates allow us to place constraints on the power of the jets and the amount of fuel required to produce such outburst. To compute the energetics of the radio source, we consider standard synchrotron minimum-energy arguments.
Specifically, using the total 1.4 GHz flux density (23.5~mJy) and source size (\(11\times3\) pc), one obtains a minimum (equipartition) energy (e.g., \citealt{govoni2004,beck2005}) of:
\begin{equation}
E_{\min} \simeq (3\mbox{--}5)\times10^{50}\ \mathrm{erg}\,,
\end{equation}
where the range corresponds to the proton-to-electron energy ratio \(k=0\mbox{--}1\). This value should be regarded as a lower limit, since the source is unlikely to be exactly at minimum energy. Importantly, \(E_{\min}\) is \emph{not} the overall kinetic budget of the outburst that is powering the jets; rather, it represents the energy deposited so far by the AGN into relativistic particles and magnetic fields that contributes to the observed synchrotron luminosity. 
We also consider that the dynamical age of the radio source on pc scales is
$t_{dyn}\approx 180\, \mathrm{yr}$
 (see Section~\ref{subsub:propermotion} and Equation~\ref{eq:tdyn}).
The corresponding (lower-limit) jet power (energy stored in the relativistic particles and magnetic fields over the source's age) can then be estimated as:
\begin{equation}
P \geq \frac{E_{\min}}{t_{dyn}} \approx (5.3\mbox{--}8.8)\times10^{40}\ \mathrm{erg\ s^{-1}}\,\,.
\end{equation}
 This is roughly 10\% of the jet power inferred for the outburst that excavated the kpc-scale X-ray cavities in the hot atmosphere \citep{david2009}, and within the range of jet powers measured for the central elliptical galaxies of groups (e.g., \citealt{eckert2021}).
If the same jet power were sustained over a typical outburst duration of \(t_{\rm tot}\sim 10\ \mathrm{Myr}\), the total injected energy would be
$E_{\rm tot} \approx 2.2\times10^{55}\ \mathrm{erg}$. Assuming a mass-to-energy conversion efficiency of \(\epsilon = 10\%\), the required fueling mass is
\begin{equation}
M = \frac{E_{\rm tot}}{\epsilon c^{2}} 
\simeq 94 - 160 \ M_{\odot}\,.
\end{equation}
Thus, a 10 Myr outburst at the inferred jet power would require $\approx10^{2}\ M_{\odot}$ of accreted mass. This is about 10\% of the mass of the molecular gas clouds detected in absorption from ALMA observations \citep{schellenberger2020}, indicating that the observed cold gas reservoir could easily fuel and sustain the jet activity. In the case where the entire molecular cloud of $M_{mol}^{cloud}\sim7\times10^{3}$~M$_{\odot}$ \citep{schellenberger2020} was consumed to power such an outburst of 10~Myr duration, the mass-to-energy conversion efficiency would be $\sim0.2\%$.


\section{Summary and conclusions}\label{sec:conc}
We have presented new VLBA observations of the central AGN in NGC 5044, combining 1.4 GHz (L band), 4.9 GHz (C band), and 8.4 GHz (X band) continuum and spectral-line datasets obtained between 2020 and 2024. These observations delivered a parsec-scale view of jet evolution and cold atomic gas in a cool-core galaxy group. Our primary findings are summarized as follows:
\begin{itemize}
    \item The 1.4 GHz VLBA image recovers a symmetric northeast–southwest jet system with a largest linear extent of $\sim$11 pc. Its position angle is consistent with previous VLBI images, and confirms a $\approx90^{\circ}$ misalignment of the $\sim 11$~pc scale jet (see also \citealt{ubertosi2024a}) with the kpc-scale radio and X-ray outburst axis (see Figure~\ref{fig:continuum} and Section \ref{subsec:1.4continuum}). In addition, naturally weighted 1.4 GHz maps reveal low surface-brightness extensions aligned with the northwest-southeast fossil outburst, possibly representing relic pc-scale plasma associated with past jet episodes.
    \item Between 2020 and 2024, the radio core decreased in flux density by 20\% at both 4.9 GHz and 8.4 GHz (Section \ref{subsub:continuumCX}). This level of variability, consistent with sub-mm variability reported previously, indicates ongoing, rapid changes in accretion and jet-launching conditions on sub-year to few-year timescales. 
    \item We detect outward motion of  multiple jet knots with an average apparent expansion speed $\langle v_{j}\rangle=(0.10\pm0.02)c$, consistent with mildly relativistic jets, based on model-fitting of the VLBA visibilities from the two epochs (see Figure \ref{fig:CXdiff}, Table \ref{tab:difmap}, and Section \ref{subsub:propermotion}). 
    \item We detect a narrow, redshifted HI absorption feature at $264~\mathrm{km\,s^{-1}}$ against the VLBA core, corresponding to a compact atomic cloud with $\mathrm{FWHM} \simeq 1.7~\mathrm{km\,s^{-1}}$ and high opacity (Figure \ref{fig:HIcore} \& \ref{fig:HIspectra}, Section \ref{subsec:spectrallineanalysis}). This line has a velocity counterpart in both MeerKAT HI and ALMA CO(2-1) data, strongly indicating that the VLBA-detected absorber traces the same multiphase cloud complex previously identified at lower spatial resolution (Figure \ref{fig:HIandCO} \& \ref{fig:absorption_model}). This close velocity correspondence between CO and HI absorption features in ALMA, MeerKAT, and the new VLBA data demonstrates that cold atomic and molecular gas coexist in infalling clouds at parsec scales.
    \item Overall, the combination of infalling $\sim$10$^{3}\, M_{\odot}$ cold clouds, mildly relativistic pc-scale jet expansion, multi-frequency core variability, and misaligned fossil outbursts provides strong evidence that the AGN in NGC~5044 cycles between inflow of cold gas clouds and repeated jet triggering on kyr - Myr timescales. 
\end{itemize}
In the future, expanding the VLBI monitoring of proper motions in NGC~5044 (over a similar 4-yr time span) would improve the estimates of jet trajectories, and potentially enable the detection of acceleration or deceleration in the jet velocity and width radial profiles. Moreover, our HI detection in NGC~5044 demonstrates that VLBI spectral-line absorption studies against faint ($\leq$10~mJy) cores are feasible, and opens a window on a broader investigation of atomic gas in central galaxies.

\begin{acknowledgments}
The authors thank the referee for their constructive and useful revision of this manuscript, that improved the analysis and presentation of the results. F. Ubertosi thanks the Smithsonian Astrophysical Observatory for the hospitality and support during his visits in 2024 and 2025, during which this project was developed and finalized, respectively. F. Ubertosi acknowledges support from the research project PRIN 2022 ``AGN-sCAN: zooming-in on the AGN-galaxy connection since the cosmic noon", contract 2022JZJBHM\_002 -- CUP J53D23001610006. W. Forman acknowledges support from NASA grants GO1-22132X, GO3-24095X, 80NSSC19K0116, the Smithsonian Institution, and the Chandra High Resolution Camera Project through NASA contract NAS8-03060. PK acknowledges the support of the Department of Atomic Energy, Government of India, under the project 12-R\&D-TFR-5.02-0700. The National Radio Astronomy Observatory is a facility of the National Science Foundation operated under cooperative agreement by Associated Universities, Inc. Basic research in radio astronomy at the Naval Research Laboratory is supported by 6.1 Base funding. 
\end{acknowledgments}

\begin{contribution}

FU, GS and EOS developed the research concept and collected the data at the basis of this work. FU analyzed the data, and was responsible for writing and submitting the manuscript. GS, EOS, JV, LPD, SG, WF provided extensive support on the interpretation of the data. All co-authors contributed equally to the review of this manuscript.

\end{contribution}

%
\facilities{NRAO}






\bibliography{sample701}{}

@ARTICLE{tamhane2026,
       author = {{Tamhane}, Prathamesh and {Sun}, Ming and {Waldron}, William and {Hosogi}, Kokoro and {da Silva}, Patricia and {Le}, Huan and {Gaspari}, Massimo and {Combes}, Francoise and {Werner}, Norbert and {Schellenberger}, Gerrit and {Fabian}, Andrew and {Canning}, Rebecca and {David}, Laurence and {Donahue}, Megan and {Voit}, Mark},
        title = "{HST view of NGC 5044: Constraints on Filament Widths, Magnetic Support, Multiphase Structure, and Comparison with Cluster Environments}",
      journal = {arXiv e-prints},
         year = 2026,
        month = feb,
          eid = {arXiv:2603.00249},
        pages = {arXiv:2603.00249},
          doi = {10.48550/arXiv.2603.00249},
archivePrefix = {arXiv},
       eprint = {2603.00249},
 primaryClass = {astro-ph.GA},
       adsurl = {https://ui.adsabs.harvard.edu/abs/2026arXiv260300249T}
}

@ARTICLE{asadanakamura2012,
       author = {{Asada}, Keiichi and {Nakamura}, Masanori},
        title = "{The Structure of the M87 Jet: A Transition from Parabolic to Conical Streamlines}",
      journal = {\apjl},
         year = 2012,
        month = feb,
       volume = {745},
       number = {2},
          eid = {L28},
        pages = {L28},
          doi = {10.1088/2041-8205/745/2/L28},
archivePrefix = {arXiv},
       eprint = {1110.1793},
 primaryClass = {astro-ph.HE},
       adsurl = {https://ui.adsabs.harvard.edu/abs/2012ApJ...745L..28A}
}

@PHDTHESIS{briggs1995,
       author = {{Briggs}, Daniel Shenon},
        title = "{High fidelity deconvolution of moderately resolved sources}",
       school = {New Mexico Institute of Mining and Technology},
         year = 1995,
        month = jan,
       adsurl = {https://ui.adsabs.harvard.edu/abs/1995PhDT.......238B}
}

@article{Kharb_2006,
	author = {Kharb, P. and O'Dea, C. P. and Baum, S. A. and Colbert, E. J. M. and Xu, C.},
	doi = {10.1086/507945},
	journal = {The Astrophysical Journal},
	month = {nov},
	number = {1},
	pages = {177},
	title = {A Radio Study of the Seyfert Galaxy Markarian 6: Implications for Seyfert Life Cycles},
	url = {https://doi.org/10.1086/507945},
	volume = {652},
	year = {2006}
}

@INPROCEEDINGS{aips1990,
       author = {{Greisen}, E.~W.},
        title = "{The Astronomical Image Processing System.}",
    booktitle = {Acquisition, Processing and Archiving of Astronomical Images},
         year = 1990,
       editor = {{Longo}, Guiseppe and {Sedmak}, Giorgio},
        month = jan,
        pages = {125-142},
       adsurl = {https://ui.adsabs.harvard.edu/abs/1990apaa.conf..125G}
}

@ARTICLE{beck2005,
       author = {{Beck}, R. and {Krause}, M.},
        title = "{Revised equipartition and minimum energy formula for magnetic field strength estimates from radio synchrotron observations}",
      journal = {Astronomische Nachrichten},
         year = 2005,
        month = jul,
       volume = {326},
       number = {6},
        pages = {414-427},
          doi = {10.1002/asna.200510366},
archivePrefix = {arXiv},
       eprint = {astro-ph/0507367},
 primaryClass = {astro-ph},
       adsurl = {https://ui.adsabs.harvard.edu/abs/2005AN....326..414B}
}

@ARTICLE{boccardi2019,
       author = {{Boccardi}, B. and {Migliori}, G. and {Grandi}, P. and {Torresi}, E. and {Mertens}, F. and {Karamanavis}, V. and {Angioni}, R. and {Vignali}, C.},
        title = "{The TeV-emitting radio galaxy 3C 264. VLBI kinematics and SED modeling}",
      journal = {\aap},
         year = 2019,
        month = jul,
       volume = {627},
          eid = {A89},
        pages = {A89},
          doi = {10.1051/0004-6361/201935183},
archivePrefix = {arXiv},
       eprint = {1905.06634},
 primaryClass = {astro-ph.HE},
       adsurl = {https://ui.adsabs.harvard.edu/abs/2019A&A...627A..89B}
}

@ARTICLE{boccardi2021,
       author = {{Boccardi}, B. and {Perucho}, M. and {Casadio}, C. and {Grandi}, P. and {Macconi}, D. and {Torresi}, E. and {Pellegrini}, S. and {Krichbaum}, T.~P. and {Kadler}, M. and {Giovannini}, G. and {Karamanavis}, V. and {Ricci}, L. and {Madika}, E. and {Bach}, U. and {Ros}, E. and {Giroletti}, M. and {Zensus}, J.~A.},
        title = "{Jet collimation in NGC 315 and other nearby AGN}",
      journal = {\aap},
         year = 2021,
        month = mar,
       volume = {647},
          eid = {A67},
        pages = {A67},
          doi = {10.1051/0004-6361/202039612},
archivePrefix = {arXiv},
       eprint = {2012.14831},
 primaryClass = {astro-ph.HE},
       adsurl = {https://ui.adsabs.harvard.edu/abs/2021A&A...647A..67B}
}

@ARTICLE{boccardi2021_proceeding,
       author = {{Boccardi}, Bia and {Madika}, Eftychia and {Ricci}, Luca},
        title = "{Accretion mode and jet collimation in active galactic nuclei}",
      journal = {Astronomische Nachrichten},
         year = 2021,
        month = nov,
       volume = {342},
       number = {1071},
        pages = {1071-1076},
          doi = {10.1002/asna.20210085},
       adsurl = {https://ui.adsabs.harvard.edu/abs/2021AN....342.1071B}
}

@software{carta2021,
       author = {{Comrie}, Angus and {Wang}, Kuo-Song and {Hsu}, Shou-Chieh and {Moraghan}, Anthony and {Harris}, Pamela and {Pang}, Qi and {Pi{\r{A}}ska}, Adrianna and {Chiang}, Cheng-Chin and {Simmonds}, Rob and {Chang}, Tien-Hao and {Jan}, Hengtai and {Lin}, Ming-Yi},
        title = "{CARTA: Cube Analysis and Rendering Tool for Astronomy}",
 howpublished = {Astrophysics Source Code Library, record ascl:2103.031},
         year = 2021,
        month = mar,
          eid = {ascl:2103.031},
archivePrefix = {ascl},
       eprint = {2103.031},
       adsurl = {https://ui.adsabs.harvard.edu/abs/2021ascl.soft03031C}
}

@ARTICLE{casa2022,
       author = {{CASA Team} and {Bean}, Ben and {Bhatnagar}, Sanjay and {Castro}, Sandra and {Donovan Meyer}, Jennifer and {Emonts}, Bjorn and {Garcia}, Enrique and {Garwood}, Robert and {Golap}, Kumar and {Gonzalez Villalba}, Justo and {Harris}, Pamela and {Hayashi}, Yohei and {Hoskins}, Josh and {Hsieh}, Mingyu and {Jagannathan}, Preshanth and {Kawasaki}, Wataru and {Keimpema}, Aard and {Kettenis}, Mark and {Lopez}, Jorge and {Marvil}, Joshua and {Masters}, Joseph and {McNichols}, Andrew and {Mehringer}, David and {Miel}, Renaud and {Moellenbrock}, George and {Montesino}, Federico and {Nakazato}, Takeshi and {Ott}, Juergen and {Petry}, Dirk and {Pokorny}, Martin and {Raba}, Ryan and {Rau}, Urvashi and {Schiebel}, Darrell and {Schweighart}, Neal and {Sekhar}, Srikrishna and {Shimada}, Kazuhiko and {Small}, Des and {Steeb}, Jan-Willem and {Sugimoto}, Kanako and {Suoranta}, Ville and {Tsutsumi}, Takahiro and {van Bemmel}, Ilse M. and {Verkouter}, Marjolein and {Wells}, Akeem and {Xiong}, Wei and {Szomoru}, Arpad and {Griffith}, Morgan and {Glendenning}, Brian and {Kern}, Jeff},
        title = "{CASA, the Common Astronomy Software Applications for Radio Astronomy}",
      journal = {\pasp},
         year = 2022,
        month = nov,
       volume = {134},
       number = {1041},
          eid = {114501},
        pages = {114501},
          doi = {10.1088/1538-3873/ac9642},
archivePrefix = {arXiv},
       eprint = {2210.02276},
 primaryClass = {astro-ph.IM},
       adsurl = {https://ui.adsabs.harvard.edu/abs/2022PASP..134k4501C}
}

@ARTICLE{david2009,
       author = {{David}, Laurence P. and {Jones}, Christine and {Forman}, William and {Nulsen}, Paul and {Vrtilek}, Jan and {O'Sullivan}, Ewan and {Giacintucci}, Simona and {Raychaudhury}, Somak},
        title = "{Isotropic Active Galactic Nucleus Heating with Small Radio-quiet Bubbles in the NGC 5044 Group}",
      journal = {\apj},
         year = 2009,
        month = nov,
       volume = {705},
       number = {1},
        pages = {624-638},
          doi = {10.1088/0004-637X/705/1/624},
archivePrefix = {arXiv},
       eprint = {0905.0654},
 primaryClass = {astro-ph.CO},
       adsurl = {https://ui.adsabs.harvard.edu/abs/2009ApJ...705..624D}
}

@ARTICLE{david2017,
       author = {{David}, Laurence P. and {Vrtilek}, Jan and {O'Sullivan}, Ewan and {Jones}, Christine and {Forman}, William and {Sun}, Ming},
        title = "{The Presence of Thermally Unstable X-Ray Filaments and the Production of Cold Gas in the NGC 5044 Group}",
      journal = {\apj},
         year = 2017,
        month = jun,
       volume = {842},
       number = {2},
          eid = {84},
        pages = {84},
          doi = {10.3847/1538-4357/aa756c},
archivePrefix = {arXiv},
       eprint = {1706.02956},
 primaryClass = {astro-ph.GA},
       adsurl = {https://ui.adsabs.harvard.edu/abs/2017ApJ...842...84D}
}

@ARTICLE{david2014,
       author = {{David}, Laurence P. and {Lim}, Jeremy and {Forman}, William and {Vrtilek}, Jan and {Combes}, Francoise and {Salome}, Philippe and {Edge}, Alastair and {Hamer}, Stephen and {Jones}, Christine and {Sun}, Ming and {O'Sullivan}, Ewan and {Gastaldello}, Fabio and {Bardelli}, Sandro and {Temi}, Pasquale and {Schmitt}, Henrique and {Ohyama}, Youichi and {Mathews}, William and {Brighenti}, Fabrizio and {Giacintucci}, Simona and {Trung}, Dinh-V.},
        title = "{Molecular Gas in the X-Ray Bright Group NGC 5044 as Revealed by ALMA}",
      journal = {\apj},
         year = 2014,
        month = sep,
       volume = {792},
       number = {2},
          eid = {94},
        pages = {94},
          doi = {10.1088/0004-637X/792/2/94},
archivePrefix = {arXiv},
       eprint = {1407.3235},
 primaryClass = {astro-ph.GA},
       adsurl = {https://ui.adsabs.harvard.edu/abs/2014ApJ...792...94D}
}

@ARTICLE{deller2018,
       author = {{Deller}, A.~T. and {Weisberg}, J.~M. and {Nice}, D.~J. and {Chatterjee}, S.},
        title = "{A VLBI Distance and Transverse Velocity for PSR B1913+16}",
      journal = {\apj},
         year = 2018,
        month = aug,
       volume = {862},
       number = {2},
          eid = {139},
        pages = {139},
          doi = {10.3847/1538-4357/aacf95},
archivePrefix = {arXiv},
       eprint = {1806.10265},
 primaryClass = {astro-ph.SR},
       adsurl = {https://ui.adsabs.harvard.edu/abs/2018ApJ...862..139D}
}

@ARTICLE{duchesne2024,
       author = {{Duchesne}, S.~W. and {Grundy}, J.~A. and {Heald}, George H. and {Lenc}, Emil and {Leung}, James K. and {McConnell}, David and {Murphy}, Tara and {Pritchard}, Joshua and {Rose}, Kovi and {Thomson}, Alec J.~M. and {Wang}, Yuanming and {Wang}, Ziteng and {Whiting}, Matthew T.},
        title = "{The Rapid ASKAP Continuum Survey V: Cataloguing the sky at 1 367.5 MHz and the second data release of RACS-mid}",
      journal = {\pasa},
         year = 2024,
        month = jan,
       volume = {41},
          eid = {e003},
        pages = {e003},
          doi = {10.1017/pasa.2023.60},
archivePrefix = {arXiv},
       eprint = {2311.12369},
 primaryClass = {astro-ph.GA},
       adsurl = {https://ui.adsabs.harvard.edu/abs/2024PASA...41....3D}
}

@ARTICLE{temi2026,
       author = {{Temi}, Pasquale and {Ubertosi}, Francesco and {Brighenti}, Fabrizio and {Maragkoudakis}, Alexandros and {Olivares}, Valeria and {Amblard}, Alexandre and {Gaspari}, Massimo and {Gitti}, Myriam and {Marcum}, Pamela M. and {Fogarty}, Kevin and {Borlaff}, Alejandro S. and {Mathews}, William G.},
        title = "{Active Galactic Nucleus Feedback and the Development of Dusty Multiphase Gas in X-Ray Emitting Elliptical Galaxies}",
      journal = {\apj},
         year = 2026,
        month = mar,
       volume = {1000},
       number = {1},
          eid = {144},
        pages = {144},
          doi = {10.3847/1538-4357/ae4972},
archivePrefix = {arXiv},
       eprint = {2602.22415},
 primaryClass = {astro-ph.GA},
       adsurl = {https://ui.adsabs.harvard.edu/abs/2026ApJ..1000..144T}
}

@ARTICLE{Grossova2022,
       author = {{Grossov{\'a}}, Romana and {Werner}, Norbert and {Massaro}, Francesco and {Lakhchaura}, Kiran and {Pl{\v{s}}ek}, Tom{\'a}{\v{s}} and {Gab{\'a}nyi}, Krisztina and {Rajpurohit}, Kamlesh and {Canning}, Rebecca E.~A. and {Nulsen}, Paul and {O'Sullivan}, Ewan and {Allen}, Steven W. and {Fabian}, Andrew},
        title = "{Very Large Array Radio Study of a Sample of Nearby X-Ray and Optically Bright Early-type Galaxies}",
      journal = {\apjs},
         year = 2022,
        month = feb,
       volume = {258},
       number = {2},
          eid = {30},
        pages = {30},
          doi = {10.3847/1538-4365/ac366c},
archivePrefix = {arXiv},
       eprint = {2111.02430},
 primaryClass = {astro-ph.HE},
       adsurl = {https://ui.adsabs.harvard.edu/abs/2022ApJS..258...30G}
}

@INPROCEEDINGS{difmap1997,
       author = {{Shepherd}, M.~C.},
        title = "{Difmap: an Interactive Program for Synthesis Imaging}",
    booktitle = {Astronomical Data Analysis Software and Systems VI},
         year = 1997,
       editor = {{Hunt}, Gareth and {Payne}, Harry},
       series = {Astronomical Society of the Pacific Conference Series},
       volume = {125},
        month = jan,
        pages = {77},
       adsurl = {https://ui.adsabs.harvard.edu/abs/1997ASPC..125...77S}
}

@ARTICLE{diniz2017,
       author = {{Diniz}, Suzi I.~F. and {Pastoriza}, Miriani G. and {Hernandez-Jimenez}, Jose A. and {Riffel}, Rogerio and {Ricci}, Tiago V. and {Steiner}, Jo{\~a}o E. and {Riffel}, Rogemar A.},
        title = "{Integral field spectroscopy of the inner kpc of the elliptical galaxy NGC 5044}",
      journal = {\mnras},
         year = 2017,
        month = sep,
       volume = {470},
       number = {2},
        pages = {1703-1717},
          doi = {10.1093/mnras/stx1322},
archivePrefix = {arXiv},
       eprint = {1705.08874},
 primaryClass = {astro-ph.GA},
       adsurl = {https://ui.adsabs.harvard.edu/abs/2017MNRAS.470.1703D}
}

@ARTICLE{donahue2022,
       author = {{Donahue}, Megan and {Voit}, G. Mark},
        title = "{Baryon cycles in the biggest galaxies}",
      journal = {\physrep},
         year = 2022,
        month = aug,
       volume = {973},
        pages = {1-109},
          doi = {10.1016/j.physrep.2022.04.005},
archivePrefix = {arXiv},
       eprint = {2204.08099},
 primaryClass = {astro-ph.GA},
       adsurl = {https://ui.adsabs.harvard.edu/abs/2022PhR...973....1D}
}

@ARTICLE{eckert2021,
       author = {{Eckert}, Dominique and {Gaspari}, Massimo and {Gastaldello}, Fabio and {Le Brun}, Amandine M.~C. and {O'Sullivan}, Ewan},
        title = "{Feedback from Active Galactic Nuclei in Galaxy Groups}",
      journal = {Universe},
         year = 2021,
        month = may,
       volume = {7},
       number = {5},
          eid = {142},
        pages = {142},
          doi = {10.3390/universe7050142},
archivePrefix = {arXiv},
       eprint = {2106.13259},
 primaryClass = {astro-ph.GA},
       adsurl = {https://ui.adsabs.harvard.edu/abs/2021Univ....7..142E}
}

@ARTICLE{gastaldello2009,
       author = {{Gastaldello}, Fabio and {Buote}, David A. and {Temi}, Pasquale and {Brighenti}, Fabrizio and {Mathews}, William G. and {Ettori}, Stefano},
        title = "{X-Ray Cavities, Filaments, and Cold Fronts in the Core of the Galaxy Group NGC 5044}",
      journal = {\apj},
         year = 2009,
        month = mar,
       volume = {693},
       number = {1},
        pages = {43-55},
          doi = {10.1088/0004-637X/693/1/43},
archivePrefix = {arXiv},
       eprint = {0807.3526},
 primaryClass = {astro-ph},
       adsurl = {https://ui.adsabs.harvard.edu/abs/2009ApJ...693...43G}
}

@ARTICLE{giacintucci2011,
       author = {{Giacintucci}, Simona and {O'Sullivan}, Ewan and {Vrtilek}, Jan and {David}, Laurence P. and {Raychaudhury}, Somak and {Venturi}, Tiziana and {Athreya}, Ramana M. and {Clarke}, Tracy E. and {Murgia}, Matteo and {Mazzotta}, Pasquale and {Gitti}, Myriam and {Ponman}, Trevor and {Ishwara-Chandra}, C.~H. and {Jones}, Christine and {Forman}, William R.},
        title = "{A Combined Low-radio Frequency/X-ray Study of Galaxy Groups. I. Giant Metrewave Radio Telescope Observations at 235 MHz AND 610 MHz}",
      journal = {\apj},
         year = 2011,
        month = may,
       volume = {732},
       number = {2},
          eid = {95},
        pages = {95},
          doi = {10.1088/0004-637X/732/2/95},
archivePrefix = {arXiv},
       eprint = {1103.1364},
 primaryClass = {astro-ph.CO},
       adsurl = {https://ui.adsabs.harvard.edu/abs/2011ApJ...732...95G}
}

@ARTICLE{giroletti2003,
       author = {{Giroletti}, M. and {Giovannini}, G. and {Taylor}, G.~B. and {Conway}, J.~E. and {Lara}, L. and {Venturi}, T.},
        title = "{Lobe advance velocities in the extragalactic compact symmetric object <ASTROBJ>4C 31.04</ASTROBJ>}",
      journal = {\aap},
         year = 2003,
        month = mar,
       volume = {399},
        pages = {889-897},
          doi = {10.1051/0004-6361:20021821},
archivePrefix = {arXiv},
       eprint = {astro-ph/0212232},
 primaryClass = {astro-ph},
       adsurl = {https://ui.adsabs.harvard.edu/abs/2003A&A...399..889G}
}

@ARTICLE{giroletti2005,
       author = {{Giroletti}, M. and {Taylor}, G.~B. and {Giovannini}, G.},
        title = "{The Two-sided Parsec-Scale Structure of the Low-Luminosity Active Galactic Nucleus in NGC 4278}",
      journal = {\apj},
         year = 2005,
        month = mar,
       volume = {622},
       number = {1},
        pages = {178-186},
          doi = {10.1086/427898},
archivePrefix = {arXiv},
       eprint = {astro-ph/0412204},
 primaryClass = {astro-ph},
       adsurl = {https://ui.adsabs.harvard.edu/abs/2005ApJ...622..178G}
}

@ARTICLE{giovannini2018,
       author = {{Giovannini}, G. and {Savolainen}, T. and {Orienti}, M. and {Nakamura}, M. and {Nagai}, H. and {Kino}, M. and {Giroletti}, M. and {Hada}, K. and {Bruni}, G. and {Kovalev}, Y.~Y. and {Anderson}, J.~M. and {D'Ammando}, F. and {Hodgson}, J. and {Honma}, M. and {Krichbaum}, T.~P. and {Lee}, S.-S. and {Lico}, R. and {Lisakov}, M.~M. and {Lobanov}, A.~P. and {Petrov}, L. and {Sohn}, B.~W. and {Sokolovsky}, K.~V. and {Voitsik}, P.~A. and {Zensus}, J.~A. and {Tingay}, S.},
        title = "{A wide and collimated radio jet in 3C84 on the scale of a few hundred gravitational radii}",
      journal = {Nature Astronomy},
         year = 2018,
        month = apr,
       volume = {2},
        pages = {472-477},
          doi = {10.1038/s41550-018-0431-2},
archivePrefix = {arXiv},
       eprint = {1804.02198},
 primaryClass = {astro-ph.GA},
       adsurl = {https://ui.adsabs.harvard.edu/abs/2018NatAs...2..472G}
}

@ARTICLE{gitti2012,
       author = {{Gitti}, Myriam and {Brighenti}, Fabrizio and {McNamara}, Brian R.},
        title = "{Evidence for AGN Feedback in Galaxy Clusters and Groups}",
      journal = {Advances in Astronomy},
         year = 2012,
        month = jan,
       volume = {2012},
          eid = {950641},
        pages = {950641},
          doi = {10.1155/2012/950641},
archivePrefix = {arXiv},
       eprint = {1109.3334},
 primaryClass = {astro-ph.CO},
       adsurl = {https://ui.adsabs.harvard.edu/abs/2012AdAst2012E...6G}
}

@ARTICLE{govoni2004,
       author = {{Govoni}, Federica and {Feretti}, Luigina},
        title = "{Magnetic Fields in Clusters of Galaxies}",
      journal = {International Journal of Modern Physics D},
         year = 2004,
        month = jan,
       volume = {13},
       number = {8},
        pages = {1549-1594},
          doi = {10.1142/S0218271804005080},
archivePrefix = {arXiv},
       eprint = {astro-ph/0410182},
 primaryClass = {astro-ph},
       adsurl = {https://ui.adsabs.harvard.edu/abs/2004IJMPD..13.1549G}
}

@ARTICLE{hogan2015b,
       author = {{Hogan}, M.~T. and {Edge}, A.~C. and {Geach}, J.~E. and {Grainge}, K.~J.~B. and {Hlavacek-Larrondo}, J. and {Hovatta}, T. and {Karim}, A. and {McNamara}, B.~R. and {Rumsey}, C. and {Russell}, H.~R. and {Salom{\'e}}, P. and {Aller}, H.~D. and {Aller}, M.~F. and {Benford}, D.~J. and {Fabian}, A.~C. and {Readhead}, A.~C.~S. and {Sadler}, E.~M. and {Saunders}, R.~D.~E.},
        title = "{High radio-frequency properties and variability of brightest cluster galaxies}",
      journal = {\mnras},
         year = 2015,
        month = oct,
       volume = {453},
       number = {2},
        pages = {1223-1240},
          doi = {10.1093/mnras/stv1518},
archivePrefix = {arXiv},
       eprint = {1507.03022},
 primaryClass = {astro-ph.GA},
       adsurl = {https://ui.adsabs.harvard.edu/abs/2015MNRAS.453.1223H}
}

@ARTICLE{hovatta2014,
       author = {{Hovatta}, Talvikki and {Aller}, Margo F. and {Aller}, Hugh D. and {Clausen-Brown}, Eric and {Homan}, Daniel C. and {Kovalev}, Yuri Y. and {Lister}, Matthew L. and {Pushkarev}, Alexander B. and {Savolainen}, Tuomas},
        title = "{MOJAVE: Monitoring of Jets in Active Galactic Nuclei with VLBA Experiments. XI. Spectral Distributions}",
      journal = {\aj},
         year = 2014,
        month = jun,
       volume = {147},
       number = {6},
          eid = {143},
        pages = {143},
          doi = {10.1088/0004-6256/147/6/143},
archivePrefix = {arXiv},
       eprint = {1404.0014},
 primaryClass = {astro-ph.GA},
       adsurl = {https://ui.adsabs.harvard.edu/abs/2014AJ....147..143H}
}

@ARTICLE{fabian2012,
       author = {{Fabian}, A.~C.},
        title = "{Observational Evidence of Active Galactic Nuclei Feedback}",
      journal = {\araa},
         year = 2012,
        month = sep,
       volume = {50},
        pages = {455-489},
          doi = {10.1146/annurev-astro-081811-125521},
archivePrefix = {arXiv},
       eprint = {1204.4114},
 primaryClass = {astro-ph.CO},
       adsurl = {https://ui.adsabs.harvard.edu/abs/2012ARA&A..50..455F}
}

@ARTICLE{kalberla2009,
       author = {{Kalberla}, Peter M.~W. and {Kerp}, J{\"u}rgen},
        title = "{The Hi Distribution of the Milky Way}",
      journal = {\araa},
         year = 2009,
        month = sep,
       volume = {47},
       number = {1},
        pages = {27-61},
          doi = {10.1146/annurev-astro-082708-101823},
       adsurl = {https://ui.adsabs.harvard.edu/abs/2009ARA&A..47...27K}
}

@article{Rao_2023,
	author = {Rao, Vaishnav V and Kharb, P and Rubinur, K and Silpa, S and Roy, N and Sebastian, B and Singh, V and Baghel, J and Manna, S and Ishwara-Chandra, C H},
	doi = {10.1093/mnras/stad1901},
	eprint = {https://academic.oup.com/mnras/article-pdf/524/2/1615/50860694/stad1901.pdf},
	issn = {0035-8711},
	journal = {\mnras},
	month = {06},
	number = {2},
	pages = {1615-1624},
	title = {AGN feedback through multiple jet cycles in the Seyfert galaxy NGC 2639},
	url = {https://doi.org/10.1093/mnras/stad1901},
	volume = {524},
	year = {2023}
}

@ARTICLE{tonry2001,
       author = {{Tonry}, John L. and {Dressler}, Alan and {Blakeslee}, John P. and {Ajhar}, Edward A. and {Fletcher}, Andr{\'e} B. and {Luppino}, Gerard A. and {Metzger}, Mark R. and {Moore}, Christopher B.},
        title = "{The SBF Survey of Galaxy Distances. IV. SBF Magnitudes, Colors, and Distances}",
      journal = {\apj},
         year = 2001,
        month = jan,
       volume = {546},
       number = {2},
        pages = {681-693},
          doi = {10.1086/318301},
archivePrefix = {arXiv},
       eprint = {astro-ph/0011223},
 primaryClass = {astro-ph},
       adsurl = {https://ui.adsabs.harvard.edu/abs/2001ApJ...546..681T}
}

@ARTICLE{sadibekova2024,
       author = {{Sadibekova}, T. and {Arnaud}, M. and {Pratt}, G.~W. and {Tarr{\'\i}o}, P. and {Melin}, J.-B.},
        title = "{MCXC-II: Second release of the Meta-Catalogue of X-ray detected Clusters of galaxies}",
      journal = {\aap},
         year = 2024,
        month = aug,
       volume = {688},
          eid = {A187},
        pages = {A187},
          doi = {10.1051/0004-6361/202449427},
archivePrefix = {arXiv},
       eprint = {2402.01538},
 primaryClass = {astro-ph.CO},
       adsurl = {https://ui.adsabs.harvard.edu/abs/2024A&A...688A.187S}
}

@ARTICLE{Buote_2003,
       author = {{Buote}, David A. and {Lewis}, Aaron D. and {Brighenti}, Fabrizio and {Mathews}, William G.},
        title = "{XMM-Newton and Chandra Observations of the Galaxy Group NGC 5044. I. Evidence for Limited Multiphase Hot Gas}",
      journal = {\apj},
         year = 2003,
        month = sep,
       volume = {594},
       number = {2},
        pages = {741-757},
          doi = {10.1086/377094},
archivePrefix = {arXiv},
       eprint = {astro-ph/0205362},
 primaryClass = {astro-ph},
       adsurl = {https://ui.adsabs.harvard.edu/abs/2003ApJ...594..741B}
}

@INPROCEEDINGS{koribalski1996,
       author = {{Koribalski}, Barbel},
        title = "{The Relation Between Large-Scale Gas Dynamics, Nuclear Kinematics, and Activity in Spiral Galaxies}",
    booktitle = {The Minnesota Lectures on Extragalactic Neutral Hydrogen},
         year = 1996,
       editor = {{Skillman}, Evan David},
       series = {Astronomical Society of the Pacific Conference Series},
       volume = {106},
        month = jan,
        pages = {238},
       adsurl = {https://ui.adsabs.harvard.edu/abs/1996ASPC..106..238K}
}

@ARTICLE{kovalev2020,
       author = {{Kovalev}, Y.~Y. and {Pushkarev}, A.~B. and {Nokhrina}, E.~E. and {Plavin}, A.~V. and {Beskin}, V.~S. and {Chernoglazov}, A.~V. and {Lister}, M.~L. and {Savolainen}, T.},
        title = "{A transition from parabolic to conical shape as a common effect in nearby AGN jets}",
      journal = {\mnras},
         year = 2020,
        month = jul,
       volume = {495},
       number = {4},
        pages = {3576-3591},
          doi = {10.1093/mnras/staa1121},
archivePrefix = {arXiv},
       eprint = {1907.01485},
 primaryClass = {astro-ph.GA},
       adsurl = {https://ui.adsabs.harvard.edu/abs/2020MNRAS.495.3576K}
}

@ARTICLE{mcnamaranulsen2007,
       author = {{McNamara}, B.~R. and {Nulsen}, P.~E.~J.},
        title = "{Heating Hot Atmospheres with Active Galactic Nuclei}",
      journal = {\araa},
         year = 2007,
        month = sep,
       volume = {45},
       number = {1},
        pages = {117-175},
          doi = {10.1146/annurev.astro.45.051806.110625},
archivePrefix = {arXiv},
       eprint = {0709.2152},
 primaryClass = {astro-ph},
       adsurl = {https://ui.adsabs.harvard.edu/abs/2007ARA&A..45..117M}
}

@ARTICLE{mcnamaranulsen2012,
       author = {{McNamara}, B.~R. and {Nulsen}, P.~E.~J.},
        title = "{Mechanical feedback from active galactic nuclei in galaxies, groups and clusters}",
      journal = {New Journal of Physics},
         year = 2012,
        month = may,
       volume = {14},
       number = {5},
          eid = {055023},
        pages = {055023},
          doi = {10.1088/1367-2630/14/5/055023},
archivePrefix = {arXiv},
       eprint = {1204.0006},
 primaryClass = {astro-ph.CO},
       adsurl = {https://ui.adsabs.harvard.edu/abs/2012NJPh...14e5023M}
}

@ARTICLE{mendel2008,
       author = {{Mendel}, J. Trevor and {Proctor}, Robert N. and {Forbes}, Duncan A. and {Brough}, Sarah},
        title = "{The anatomy of the NGC5044 group - I. Group membership and dynamics}",
      journal = {\mnras},
         year = 2008,
        month = sep,
       volume = {389},
       number = {2},
        pages = {749-765},
          doi = {10.1111/j.1365-2966.2008.13514.x},
archivePrefix = {arXiv},
       eprint = {0806.3127},
 primaryClass = {astro-ph},
       adsurl = {https://ui.adsabs.harvard.edu/abs/2008MNRAS.389..749M}
}

@ARTICLE{mirabel1994,
       author = {{Mirabel}, I.~F. and {Rodr{\'\i}guez}, L.~F.},
        title = "{A superluminal source in the Galaxy}",
      journal = {\nat},
         year = 1994,
        month = sep,
       volume = {371},
       number = {6492},
        pages = {46-48},
          doi = {10.1038/371046a0},
       adsurl = {https://ui.adsabs.harvard.edu/abs/1994Natur.371...46M}
}

@ARTICLE{park2019,
       author = {{Park}, Jongho and {Hada}, Kazuhiro and {Kino}, Motoki and {Nakamura}, Masanori and {Hodgson}, Jeffrey and {Ro}, Hyunwook and {Cui}, Yuzhu and {Asada}, Keiichi and {Algaba}, Juan-Carlos and {Sawada-Satoh}, Satoko and {Lee}, Sang-Sung and {Cho}, Ilje and {Shen}, Zhiqiang and {Jiang}, Wu and {Trippe}, Sascha and {Niinuma}, Kotaro and {Sohn}, Bong Won and {Jung}, Taehyun and {Zhao}, Guang-Yao and {Wajima}, Kiyoaki and {Tazaki}, Fumie and {Honma}, Mareki and {An}, Tao and {Akiyama}, Kazunori and {Byun}, Do-Young and {Kim}, Jongsoo and {Zhang}, Yingkang and {Cheng}, Xiaopeng and {Kobayashi}, Hideyuki and {Shibata}, Katsunori M. and {Lee}, Jee Won and {Roh}, Duk-Gyoo and {Oh}, Se-Jin and {Yeom}, Jae-Hwan and {Jung}, Dong-Kyu and {Oh}, Chungsik and {Kim}, Hyo-Ryoung and {Hwang}, Ju-Yeon and {Hagiwara}, Yoshiaki},
        title = "{Kinematics of the M87 Jet in the Collimation Zone: Gradual Acceleration and Velocity Stratification}",
      journal = {\apj},
         year = 2019,
        month = dec,
       volume = {887},
       number = {2},
          eid = {147},
        pages = {147},
          doi = {10.3847/1538-4357/ab5584},
archivePrefix = {arXiv},
       eprint = {1911.02279},
 primaryClass = {astro-ph.HE},
       adsurl = {https://ui.adsabs.harvard.edu/abs/2019ApJ...887..147P}
}

@ARTICLE{scheuer_1979,
       author = {{Scheuer}, P.~A.~G. and {Readhead}, A.~C.~S.},
        title = "{Superluminally expanding radio sources and the radio-quiet QSOs}",
      journal = {\nat},
         year = 1979,
        month = jan,
       volume = {277},
        pages = {182-185},
          doi = {10.1038/277182a0},
       adsurl = {https://ui.adsabs.harvard.edu/abs/1979Natur.277..182S}
}

@ARTICLE{rajpurohit2025,
       author = {{Rajpurohit}, Kamlesh and {Deb}, Tirna and {Kolokythas}, Konstantinos and {Thorat}, Kshitij and {O'Sullivan}, Ewan and {Schellenberger}, Gerrit and {David}, Laurence P. and {Vrtilek}, Jan M. and {Giacintucci}, Simona and {Forman}, William and {Jones}, Christine and {Ramatsoku}, Mpati},
        title = "{Revisiting the Group-dominant Elliptical NGC 5044 in the Radio Band: Continuum Emission and Detection of H I Absorption}",
      journal = {\apj},
         year = 2025,
        month = may,
       volume = {984},
       number = {2},
          eid = {120},
        pages = {120},
          doi = {10.3847/1538-4357/adc1cb},
archivePrefix = {arXiv},
       eprint = {2501.02076},
 primaryClass = {astro-ph.GA},
       adsurl = {https://ui.adsabs.harvard.edu/abs/2025ApJ...984..120R}
}

@ARTICLE{rose2019,
       author = {{Rose}, Tom and {Edge}, A.~C. and {Combes}, F. and {Gaspari}, M. and {Hamer}, S. and {Nesvadba}, N. and {Russell}, H. and {Tremblay}, G.~R. and {Baum}, S.~A. and {O'Dea}, C. and {Peck}, A.~B. and {Sarazin}, C. and {Vantyghem}, A. and {Bremer}, M. and {Donahue}, M. and {Fabian}, A.~C. and {Ferland}, G. and {McNamara}, B.~R. and {Mittal}, R. and {Oonk}, J.~B.~R. and {Salom{\'e}}, P. and {Swinbank}, A.~M. and {Voit}, M.},
        title = "{Deep and narrow CO absorption revealing molecular clouds in the Hydra-A brightest cluster galaxy}",
      journal = {\mnras},
         year = 2019,
        month = may,
       volume = {485},
       number = {1},
        pages = {229-238},
          doi = {10.1093/mnras/stz406},
archivePrefix = {arXiv},
       eprint = {1902.01863},
 primaryClass = {astro-ph.GA},
       adsurl = {https://ui.adsabs.harvard.edu/abs/2019MNRAS.485..229R}
}

@ARTICLE{rose2022,
       author = {{Rose}, Tom and {Edge}, Alastair and {Kiehlmann}, Sebastian and {Baek}, Junhyun and {Chung}, Aeree and {Jung}, Tae-Hyun and {Kim}, Jae-Woo and {Readhead}, Anthony C.~S. and {Sedgewick}, Aidan},
        title = "{The variability of brightest cluster galaxies at high radio frequencies}",
      journal = {\mnras},
         year = 2022,
        month = jan,
       volume = {509},
       number = {2},
        pages = {2869-2884},
          doi = {10.1093/mnras/stab3217},
archivePrefix = {arXiv},
       eprint = {2109.03241},
 primaryClass = {astro-ph.GA},
       adsurl = {https://ui.adsabs.harvard.edu/abs/2022MNRAS.509.2869R}
}

@ARTICLE{rose2023,
       author = {{Rose}, Tom and {McNamara}, B.~R. and {Combes}, F. and {Edge}, A.~C. and {Fabian}, A.~C. and {Gaspari}, M. and {Russell}, H. and {Salom{\'e}}, P. and {Tremblay}, G. and {Ferland}, G.},
        title = "{Does absorption against AGN reveal supermassive black hole accretion?}",
      journal = {\mnras},
         year = 2023,
        month = jan,
       volume = {518},
       number = {1},
        pages = {878-892},
          doi = {10.1093/mnras/stac3194},
archivePrefix = {arXiv},
       eprint = {2210.14922},
 primaryClass = {astro-ph.GA},
       adsurl = {https://ui.adsabs.harvard.edu/abs/2023MNRAS.518..878R}
}

@ARTICLE{schellenberger2020,
       author = {{Schellenberger}, Gerrit and {David}, Laurence P. and {Vrtilek}, Jan and {O'Sullivan}, Ewan and {Lim}, Jeremy and {Forman}, William and {Sun}, Ming and {Combes}, Francoise and {Salome}, Philippe and {Jones}, Christine and {Giacintucci}, Simona and {Edge}, Alastair and {Gastaldello}, Fabio and {Temi}, Pasquale and {Brighenti}, Fabrizio and {Bardelli}, Sandro},
        title = "{Atacama Compact Array Measurements of the Molecular Mass in the NGC 5044 Cooling-flow Group}",
      journal = {\apj},
         year = 2020,
        month = may,
       volume = {894},
       number = {1},
          eid = {72},
        pages = {72},
          doi = {10.3847/1538-4357/ab879c},
archivePrefix = {arXiv},
       eprint = {2004.01717},
 primaryClass = {astro-ph.GA},
       adsurl = {https://ui.adsabs.harvard.edu/abs/2020ApJ...894...72S}
}

@ARTICLE{schellenberger2021,
       author = {{Schellenberger}, Gerrit and {David}, Laurence P. and {Vrtilek}, Jan and {O'Sullivan}, Ewan and {Giacintucci}, Simona and {Forman}, William and {Jones}, Christine and {Venturi}, Tiziana},
        title = "{A New Feedback Cycle in the Archetypal Cooling Flow Group NGC 5044}",
      journal = {\apj},
         year = 2021,
        month = jan,
       volume = {906},
       number = {1},
          eid = {16},
        pages = {16},
          doi = {10.3847/1538-4357/abc488},
archivePrefix = {arXiv},
       eprint = {2010.13804},
 primaryClass = {astro-ph.GA},
       adsurl = {https://ui.adsabs.harvard.edu/abs/2021ApJ...906...16S}
}

@ARTICLE{schellenberger2024,
       author = {{Schellenberger}, Gerrit and {O'Sullivan}, Ewan and {David}, Laurence P. and {Vrtilek}, Jan and {Romero}, Charles and {Petitpas}, Glen and {Forman}, William and {Giacintucci}, Simona and {Gurwell}, Mark and {Jones}, Christine and {Rajpurohit}, Kamlesh and {Ubertosi}, Francesco and {Venturi}, Tiziana},
        title = "{Probing the High-frequency Variability of NGC 5044: The Key to Active Galactic Nucleus Feedback}",
      journal = {\apj},
         year = 2024,
        month = dec,
       volume = {976},
       number = {2},
          eid = {246},
        pages = {246},
          doi = {10.3847/1538-4357/ad89bc},
archivePrefix = {arXiv},
       eprint = {2409.06039},
 primaryClass = {astro-ph.GA},
       adsurl = {https://ui.adsabs.harvard.edu/abs/2024ApJ...976..246S}
}

@ARTICLE{solomon1987,
       author = {{Solomon}, P.~M. and {Rivolo}, A.~R. and {Barrett}, J. and {Yahil}, A.},
        title = "{Mass, Luminosity, and Line Width Relations of Galactic Molecular Clouds}",
      journal = {\apj},
         year = 1987,
        month = aug,
       volume = {319},
        pages = {730},
          doi = {10.1086/165493},
       adsurl = {https://ui.adsabs.harvard.edu/abs/1987ApJ...319..730S}
}

@ARTICLE{tremblay2016,
       author = {{Tremblay}, Grant R. and {Oonk}, J.~B. Raymond and {Combes}, Fran{\c{c}}oise and {Salom{\'e}}, Philippe and {O'Dea}, Christopher P. and {Baum}, Stefi A. and {Voit}, G. Mark and {Donahue}, Megan and {McNamara}, Brian R. and {Davis}, Timothy A. and {McDonald}, Michael A. and {Edge}, Alastair C. and {Clarke}, Tracy E. and {Galv{\'a}n-Madrid}, Roberto and {Bremer}, Malcolm N. and {Edwards}, Louise O.~V. and {Fabian}, Andrew C. and {Hamer}, Stephen and {Li}, Yuan and {Maury}, Ana{\"e}lle and {Russell}, Helen R. and {Quillen}, Alice C. and {Urry}, C. Megan and {Sanders}, Jeremy S. and {Wise}, Michael W.},
        title = "{Cold, clumpy accretion onto an active supermassive black hole}",
      journal = {\nat},
         year = 2016,
        month = jun,
       volume = {534},
       number = {7606},
        pages = {218-221},
          doi = {10.1038/nature17969},
archivePrefix = {arXiv},
       eprint = {1606.02304},
 primaryClass = {astro-ph.GA},
       adsurl = {https://ui.adsabs.harvard.edu/abs/2016Natur.534..218T}
}

@ARTICLE{ubertosi2024a,
       author = {{Ubertosi}, Francesco and {Schellenberger}, Gerrit and {O'Sullivan}, Ewan and {Vrtilek}, Jan and {Giacintucci}, Simona and {David}, Laurence P. and {Forman}, William and {Gitti}, Myriam and {Venturi}, Tiziana and {Jones}, Christine and {Brighenti}, Fabrizio},
        title = "{Jet Reorientation in Central Galaxies of Clusters and Groups: Insights from VLBA and Chandra Data}",
      journal = {\apj},
         year = 2024,
        month = jan,
       volume = {961},
       number = {1},
          eid = {134},
        pages = {134},
          doi = {10.3847/1538-4357/ad11d8},
archivePrefix = {arXiv},
       eprint = {2312.02283},
 primaryClass = {astro-ph.GA},
       adsurl = {https://ui.adsabs.harvard.edu/abs/2024ApJ...961..134U}
}

@ARTICLE{walker2018,
       author = {{Walker}, R. Craig and {Hardee}, Philip E. and {Davies}, Frederick B. and {Ly}, Chun and {Junor}, William},
        title = "{The Structure and Dynamics of the Subparsec Jet in M87 Based on 50 VLBA Observations over 17 Years at 43 GHz}",
      journal = {\apj},
         year = 2018,
        month = mar,
       volume = {855},
       number = {2},
          eid = {128},
        pages = {128},
          doi = {10.3847/1538-4357/aaafcc},
archivePrefix = {arXiv},
       eprint = {1802.06166},
 primaryClass = {astro-ph.HE},
       adsurl = {https://ui.adsabs.harvard.edu/abs/2018ApJ...855..128W}
}

@ARTICLE{wolfire2003,
       author = {{Wolfire}, Mark G. and {McKee}, Christopher F. and {Hollenbach}, David and {Tielens}, A.~G.~G.~M.},
        title = "{Neutral Atomic Phases of the Interstellar Medium in the Galaxy}",
      journal = {\apj},
         year = 2003,
        month = apr,
       volume = {587},
       number = {1},
        pages = {278-311},
          doi = {10.1086/368016},
archivePrefix = {arXiv},
       eprint = {astro-ph/0207098},
 primaryClass = {astro-ph},
       adsurl = {https://ui.adsabs.harvard.edu/abs/2003ApJ...587..278W}
}
\bibliographystyle{aasjournalv7}



\end{document}